\documentclass[journal]{IEEEtran}
\IEEEoverridecommandlockouts

\usepackage[linesnumbered]{algorithm2e}
\usepackage{enumitem}
\usepackage{color}
\usepackage{times,amsmath,epsfig}
\usepackage{amssymb}
\usepackage{subfigure}
\usepackage{cite}
\usepackage{multirow}
\usepackage{amsthm}

\usepackage{hyperref}
\usepackage{algorithmic}
\usepackage{rotating}
\usepackage{cleveref}

\usepackage{needspace}

\usepackage{tikz}
\usetikzlibrary{shapes,arrows}
\input{mysymbol.sty}
\tikzstyle{phantom vertex} = [ ellipse, 
                               anchor = center, 
                               minimum height = 1*\unit, 
                               minimum width  = 1*\unit,
                               inner sep=0pt,
                               anchor=center]
\tikzstyle{red vertex}   = [black, fill = red!20,   phantom vertex, draw]
\tikzstyle{black vertex} = [black, fill = black!20, phantom vertex, draw]
\tikzstyle{blue vertex}  = [black, fill = blue!20,  phantom vertex, draw]
\tikzstyle{green vertex} = [black, fill = green!20,  phantom vertex, draw]
\tikzstyle{yellow vertex} = [black, fill = yellow!20,  phantom vertex, draw]
\tikzstyle{cyan vertex} = [black, fill = cyan!20,  phantom vertex, draw]
\tikzstyle{vertex}       = [draw, phantom vertex]

\tikzstyle{point} = [ellipse, inner sep=0pt, draw, fill=white, anchor = center,
                     minimum height = 0.05*\unit, minimum width  = 0.05*\unit]

\newcommand{\QED}{\hfill\ensuremath{\blacksquare}}

\theoremstyle{plain}
\newtheorem{theorem}{Theorem}
\newtheorem{proposition}{Proposition}

\newtheorem{assumption}{Assumption}

\title{Polynomial Graphical Lasso: Learning Edges from Gaussian Graph-Stationary Signals}
\author{Andrei Buciulea,~\IEEEmembership{Student Member,~IEEE}, Jiaxi Ying,~\IEEEmembership{Member,~IEEE},
        Antonio~G.~Marques,~\IEEEmembership{Senior Member,~IEEE}, and Daniel P. Palomar,~\IEEEmembership{Fellow,~IEEE}
	
 \thanks{Work partially supported by the Spanish Spanish AEI Grants PID2019-105032GB-I00 and PID2022-136887NB-I00, and by the Autonomous Community of Madrid within the ELLIS Unit Madrid framework, URJC/CAM grant F861 and URJC/CAM grant PREDOC20-003. Work partially supported by the Hong Kong GRF 16206123 research grant. An early preliminary version of this work was presented as a conference paper in~\cite{buciulea2023learning}. \textit{(Corresponding author: Antonio G. Marques)}
    
Andrei Buciulea and Antonio G. Marques are with the Dept. of Signal Theory and Comms., King Juan Carlos University, 28933 Madrid, Spain (e-mail:{andrei.buciulea, antonio.garcia.marques}@urjc.es). 
    
Jiaxi Ying  is with the Department of Mathematics, Hong Kong University
of Science and Technology, Clear Water Bay, Kowloon, Hong Kong (e-mail:jx.ying@connect.ust.hk).

Daniel P. Palomar is with the Department of Electronic and Computer
Engineering and the Department of Industrial Engineering and Decision
Analytics, Hong Kong University of Science and Technology, Kowloon, Hong
Kong (e-mail: palomar@ust.hk).}}

\begin{document}
\maketitle
\begin{abstract}

This paper introduces Polynomial Graphical Lasso (PGL), a new approach to learning graph structures from nodal signals. Our key contribution lies in modeling the signals as Gaussian and stationary on the graph, enabling the development of a graph-learning formulation that combines the strengths of graphical lasso with a more encompassing model. Specifically, we assume that the precision matrix can take any polynomial form of the sought graph, allowing for increased flexibility in modeling nodal relationships. Given the resulting complexity and nonconvexity of the resulting optimization problem, we (i) propose a low-complexity algorithm that alternates between estimating the graph and precision matrices, and (ii) characterize its convergence. We evaluate the performance of PGL through comprehensive numerical simulations using both synthetic and real data, demonstrating its superiority over several alternatives. Overall, this approach presents a significant advancement in graph learning and holds promise for various applications in graph-aware signal analysis and beyond.

\end{abstract}
%

\section{Introduction}\label{S:Introduction}

Modern datasets often exhibit an irregular non-Euclidean support. In such scenarios, graphs have emerged as a pivotal tool, facilitating the generalization of classical information processing and structured learning techniques to irregular domains.
Today, there is a wide range of applications that leverage graphs when processing, learning, and extracting knowledge from their associated datasets (see, e.g., problems in the context of electrical, communication, social, geographic, financial, genetic, and brain networks~\cite{kolaczyk2009book,sporns2012book,EmergingFieldGSP,ortega_2018_graph,alglearninggraphs}, to name a few).
When using graphs to process structured non-Euclidean data, it is usually assumed that the underlying network topology is known. Unfortunately, this is not always the case. 
In many cases, the structure of the graph is not well defined, either because there is no underlying physical network or because the (best) metric to assess the level of association between the nodes is not known. 
 
Since in most cases the existing relationships are not known beforehand, the standard approach is to infer the structure of the network from a set of available nodal observations/signals/features.
To estimate the interactions between the existing nodes, the first step is to formally define the relationship between the topology of the graph and the properties of the signals defined on top of it. 
Early graph topology inference methods~\cite{mateos2019connecting,sardellitti2019graph}  adopted a statistical approach, such as the correlation network~\cite[Ch. 7.3.1]{kolaczyk2009book}, partial correlations, or Gaussian Markov random fields (GMRF), with the latter leading to the celebrated graphical Lasso (GL) scheme~\cite{meinshausen06,kolaczyk2009book}. Partial correlation methods have been generalized to nonlinear settings \cite{Karanikolas_icassp16}. Also in the nonlinear realm, less rigorous approaches simply postulate a similarity scores, with links being drawn if their score exceeds a given threshold. In recent years, graph signal processing (GSP) based models~\cite{MeiGraphStructure,egilmez2017graph,segarra2017network} have brought new ideas to the field, considering more complex relationships between the signals and the sought graph. These approaches have been generalized to deal with more complex scenarios that often arise in practice, such as the presence of hidden variables~\cite{chandrasekaran2012latent,chang2019graphical,buciulea2022learning} or the simultaneous inference of multiple networks~\cite{danaher2014joint,navarro2022joint}. 

The existing graph-learning methods exhibit different pros and cons, with relevant tradeoffs including computational complexity, expressiveness, model accuracy, or sample complexity, to name a few. For instance, correlation networks need very few samples and can be run in parallel for each pair of nodes, but fail to capture intermediation nodal effects. On the other hand, GL (a maximum likelihood estimator for GMRF) can handle the intermediation effect while still requiring a relatively small number of samples compared to the size of the network. Some disadvantages of GL include assuming a relatively simple signal model (failing to deal with, e.g., linear autoregressive network models) and forcing the learned graph to be a positive definite matrix. To overcome some of these issues, \cite{segarra2017network} proposed a more general model that assumed that the signals were stationary in the network (GSR) or, equivalently, that the covariance matrix can be defined as a polynomial of the adjacency of the graph \cite{marques2017stationary}. 
Since GSR is a more general model, it is less restrictive in terms of the signals it can handle. 
However, it has the disadvantage of requiring a significantly larger number of observations than GL \cite{segarra2017network}.

Our proposal is to combine the advantages of assuming Gaussianity, which implies solving a maximum likelihood problem that requires fewer node observations, with the larger generality of graph-stationary approaches. Our ultimate goal is to generalize the range of scenarios where GL can be used, while keeping the number of observations and computational complexity under control. 
To be more precise, we introduce Polynomial Graphical Lasso (PGL), a new scheme to learn graphs from signals that works under the assumption that the samples are Gaussian and graph stationary, so that the covariance (precision) matrix of the observations can be written as a polynomial of a sparse graph. These assumptions give rise to a constrained log-likelihood minimization that is jointly optimized over the precision and adjacency matrices, with GL being a particular instance of our problem.   
%
The price to pay is that the postulated optimization, even after relaxing the sparsity constraints, is more challenging, leading to a biconvex problem. To mitigate this issue we provide an efficient alternating algorithm with convergence guarantees.

\vspace{0.3cm}
\noindent\textbf{{Contributions.}} 
\vspace{0.0cm}
To summarize, our main contributions are:

\begin{itemize}
    \vspace{0.1cm}
    \item Introducing PGL, a novel graph-learning scheme that, by assuming that the observations are Gaussian and graph-stationary, generalizes GL and is able to learn a meaningful graph structure in scenarios where the precision/covariance matrices are polynomials of the sparse matrix that represents the graph.
    \vspace{0.1cm}
    \item Formulating the inference problem as a biconvex constrained optimization, with the variables to optimize being the precision matrix and the graph. While our focus is on learning the graphs, note that this implies that PGL can also be used in the context of covariance estimation. 
    \vspace{0.1cm}
    \item Developing an efficient algorithm to solve the proposed optimization, together with convergence guarantees to a stationary point (block coordinatewise minimizer).
    \vspace{0.1cm}
    \item Evaluating the performance of the proposed approach through comparisons with alternatives from the literature on synthetic and real-world datasets.
\end{itemize}
\vspace{0.2cm}

\noindent\textbf{{Outline.}} The remainder of the paper is organized as follows. Section~\ref{S:fundamental_GSP} surveys the basic graph and GSP background, placing special emphasis on the concepts of graph stationarity and Gaussian graph signals. In Section~\ref{S:netw_reconstruction}, the problem of learning (inferring) graphs from signals under different assumptions on the observations is formally stated. Section~\ref{sec:biconvexrelax_PGL_AlgDesign} presents a computationally tractable relaxation of the graph-learning problem, along with an efficient algorithm and its associated convergence guarantees. Section~\ref{S:numerical_experiments} quantifies and compares the recovery performance of the proposed approach with other methods from the literature using both synthetic and real data. Section~\ref{S:conclusions} closes the paper with concluding remarks.

\section{Graphs and Random Graph Signals}\label{S:fundamental_GSP}


This section introduces the notation used in the paper and fundamental concepts of graphs and GSP that will help to explain the relationship between the available signals and the topology of the underlying graph. 

\vspace{0.1cm}
\noindent\textbf{Notation.}
We represent scalar, vectors and matrices using lowercase ($x \in \reals$), bold lowercase ($\bbx \in \reals^{N}$), and bold uppercase ($\bbX \in \reals^{N \times R}$) letters, respectively. $X_{ij}$ represents the values of the entry $[i,j]$ of matrix $\bbX$. $\bbI_{N}$ denotes the identity matrix of size $N \times N$. The expression diag($\bbx$) represents a diagonal matrix with the values of $\bbx$ in the diagonal. The expressions $\|\cdot\|_0$ and $\|\cdot\|_1$ refer to the entry-wise (vectorized) $\ell_0$ and $\ell_1$ matrix norms, and , and $\|\cdot\|_F$ is the Frobenious norm.    
    
\vspace{0.1cm}
\noindent\textbf{Graphs.}
 Let $\ccalG = \{ \ccalV, \ccalE\}$ be a weighted and undirected graph with $N$ nodes, where $\ccalV$ and $\ccalE$ denote the vertex and edge set, respectively. The topology of the graph $\ccalG$ is encoded in the sparse, weighted adjacency matrix $\bbA\in\reals^{N \times N}$, where $A_{ij}$ represents the weight of the edge between nodes $j$ and $i$, and $A_{ij}=0$ if $i$ and $j$ are not connected. A more general representation of the graph is the Graph Shift Operator (GSO), denoted by $\bbS \in \mathbb{R}^{N \times N}$, where $S_{ij} \neq 0$ if and only if $i=j$ or $(i,j) \in \ccalE$. 
 Common choices for the GSO are the adjacency matrix $\bbA$ \cite{SandryMouraSPG_TSP13}, the combinatorial graph Laplacian $\bbL := \diag(\bbA \textbf{1}) - \bbA$ \cite{EmergingFieldGSP}, and their degree-normalized variants. Since the graph is undirected, the GSO is symmetric and can be diagonalized as $\bbS=\bbV\bbLambda \bbV^\top$, where the orthogonal matrix $\bbV\in\reals^{N \times N}$ contains the eigenvectors of the GSO, and the diagonal matrix $\bbLambda$ contains its eigenvalues. 

\vspace{0.1cm}
\noindent\textbf{Graph signals and filters.} A graph signal is represented by a vector $\bbx\in \mathbb{R}^{N}$, where $x_{i}$ denotes the signal value observed at node $i$. Alternatively, a graph signal can be understood as a $\ccalV \rightarrow \reals$ mapping, so that the signal values can represent features associated with the node at hand. Graph filters provide a flexible tool for either processing or modeling the relationship between a signal $\bbx$ and its underlying graph $\ccalG$. Succinctly, a graph filter is a linear operator $\bbH\in\reals^{N\times N}$ that takes into account the topology of the graph and is defined as a polynomial of the GSO $\bbS$ of the form
\begin{equation}\label{E:graph_filter}
    \bbH = \sum_{l=0}^{L-1} h_{l} \bbS^{l}= \sum_{l=0}^{L-1} h_{l}\bbV\bbLambda^l\bbV^\top=\bbV\Big(\sum_{l=0}^{L-1} h_{l}\bbLambda^l\Big)\bbV^\top,
\end{equation}
where $\{h_l\}_{l=0}^{L-1}$ represent the filter coefficients, and $\bbV$ and $\bbLambda$ are the eigenvectors and eigenvalues of the GSO respectively.
Since $\bbH$ is a polynomial of $\bbS$, it readily follows that both matrices have the same eigenvectors $\bbV$.

\vspace{1mm}
\noindent\textbf{Stationary graph signals.} 
A graph signal $\bbx$ is defined to be stationary on the graph $\ccalG$ if it can be expressed as the output of a graph filter $\bbH$ excited by a zero-mean white signal $\bbw \in \mathbb{R}^{N}$ \cite{marques2017stationary,perraudin2017stationary,Girault15}. 
In other words, if $\bbw$ has a covariance of $\mathbb{E}[\bbw\bbw^\top]=\bbI_N$, and $\bbx=\bbH\bbw$, then the covariance of $\bbx$ is $\bbSigma= \mathbb{E}[\bbx\bbx^\top]=\bbH\mathbb{E}[\bbw\bbw^\top]\bbH^\top=\bbH\bbH^\top=\bbH^2$. This reveals that assuming that $\bbx$ is stationary on $\ccalG$ is equivalent to saying that the covariance matrix $\bbSigma$ can be written as a polynomial of the GSO $\bbS$, and that, as a result, $\bbSigma$ and $\bbS$  have the same eigenvectors $\bbV$~[cf. \eqref{E:graph_filter}]. 
Hence, graph stationarity implies that the matrices $\bbS$ and $\bbSigma$ commute, which is a significant property to be exploited later on.
    
\vspace{0.1cm}
\noindent\textbf{Gaussian graph signals.}
A Gaussian graph signal is an $N$-dimensional \emph{random} vector $\bbx$ whose (vertex-indexed) entries are jointly Gaussian. Hence, if $\bbx \in \reals^N$ is distributed as $\mathcal{N}(\bb0,\bbSigma)$, we have that both the covariance matrix $\bbSigma$ and the precision matrix $\bbTheta=\bbSigma^{-1}$ are $N\times N$ matrices whose entries capture statistical relationships between the nodal values. In particular, using the distribution of the Gaussian, it holds that:

\begin{equation}\label{E:G_sig}
f_{\bbTheta}(\bbx) = (2\pi)^{-N/2} \cdot \det(\bbTheta)^{\frac{1}{2}} \cdot \exp^{-\frac{1}{2}\bbx^{T}\bbTheta\bbx},
\end{equation}
demonstrating that, in the context of graph signals, the entries of $\bbTheta\in\reals^{N\times N}$ account for conditional dependence relationships between the (features of the) nodes of the graph \cite{mateos2019connecting}. 

Suppose now that we have a collection of $R$ Gaussian signals $\bbX = [\bbx_1,...,\bbx_R] \in \mathbb{R}^{N \times R}$ each of them independently drawn from the distribution in \eqref{E:G_sig}. The log-likelihood associated with $\bbX$ is
\begin{equation}\label{E:Like}
    L(\bbX|\bbTheta) = \prod_{r=1}^{R} f_{\bbTheta}(\bbx_r),  \ \ \ \ccalL(\bbX|\bbTheta) = \sum_{r=1}^{R} \log(f_{\bbTheta}(\bbx_r)).
\end{equation}
This expression will be exploited when formulating our proposed inference approach and establishing links with classical methods.
\section{Graph learning problem formulation}\label{S:netw_reconstruction}

This section begins with a formal definition of the graph learning problem, followed by an explanation of some common approaches used in the literature to tackle this problem. 
Afterwards, we proceed to formalize the learning problem we aim to address and cast it as an optimization problem. 
We then provide an overview of the key features of our formulation and conduct a comparative analysis with the two closest approaches available in the literature.

To formally state the graph learning problem, let us recall that we assume: i) $\ccalG$ is an undirected graph with $N$ nodes, ii) there is a random process associated with $\ccalG$, and iii) we denote by $\bbX=[\bbx_1,...,\bbx_R] \in \reals^{N \times R}$ a collection of $R$ independent realizations of such a process. 
The goal in graph learning is to use a given set of nodal observations to find the (estimates of the) links/associations between the nodes in the graph (i.e., use $\bbX$ to estimate the $\bbS$ associated with $\ccalG$). 

This problem has been addressed under different approaches~\cite{GLasso2008, dong_2019_learning, marques2017stationary}. 
Differences among these models typically arise from the underlying assumptions that are made regarding 1) the graph, which almost universally entails just considering that the graph is sparse (see, e.g., \cite{sevilla2024estimation} for a recent exception), 2) the signals, which assume certain properties related to the nature of the signals such as smoothness \cite{dong_2019_learning} or Gaussianity, \cite{GLasso2008} among others, and 3) the relationship between the graph and the signal, such as the stationarity property \cite{marques2017stationary}.

The model we propose aims to incorporate several assumptions about both the graph and the graph signals. To that end, we propose an approach for which we assume that: 1) the graph is sparse, 2) the signals are Gaussian and, 3) the signals are stationary in the underlying graph. 

\noindent Having established these assumptions, we now proceed to formally state our graph learning problem as follows

\vspace{0.2cm}
\textbf{Problem 1} \textit{Given a set of signals $\bbX \in \reals^{N \times R}$, find the underlying sparse graph structure encoded in $\bbS$ under the assumptions:} 

\noindent \textit{(AS1): The graph signals $\bbX$ are i.i.d. realizations of $\ccalN(\bb0, \bbSigma)$.}

\noindent\textit{(AS2): The graph signals $\bbX$ are stationary in $\bbS$.} \\

\noindent Our approach is to recast Problem 1 as the following optimization 
\begin{alignat}{2}\label{E:alg_GGSR}
    \!\!&\!\underset{\bbTheta\succeq 0,\bbS \in \ccalS} {\mathsf{minimize}} \ \nonumber
    && -\log(\det(\bbTheta)) + \tr(\hbSigma\bbTheta) + \rho \|\bbS\|_0.\\
        \!\!&\!\mathsf{subject~to} && \;\; \bbTheta\bbS=\bbS\bbTheta,
\end{alignat}
where $\ccalS$ is a generic set representing additional constraints that $\bbS$ is known to satisfy (e.g., the GSO being symmetric and its entries being between zero and one). The minimization takes as input the sample covariance matrix $\hbSigma=\frac{1}{R}\bbX\bbX^\mathsf{T}$ and generates as output the estimate for $\bbS$ and, as a byproduct, the estimate for $\bbTheta$. For the problem in \eqref{E:alg_GGSR}, we require $\bbTheta$ to be positive semidefinite. This constraint arises because $\bbTheta$ is the inverse of $\hbSigma$ which is symmetric and positive definite by construction. Consequently, $\bbTheta$ inherits the properties of being symmetric and positive semidefinite.

Next, we explain the motivation for each term in \eqref{E:alg_GGSR} with special emphasis on the constraint $\bbTheta\bbS=\bbS\bbTheta$, which is a fundamental component of our approach. 
\begin{itemize}
\item 
The first two terms in the objective function are due to (AS1) and arise from minimizing the negative log-likelihood expression in \eqref{E:Like}. Indeed, it is clear that substituting \eqref{E:G_sig} into \eqref{E:Like} yields 
\begin{align*}
 \sum_{r=1}^R\left(-\frac{N}{2}\log(2\pi)-\log(\det(\bbTheta)) + \tr(\bbx_r^T\bbTheta\bbx_r)\right).   
\end{align*}
Since constants are irrelevant for the optimization, we drop the first term and divide the other two by $R$, yielding 
\begin{align}
    &-\log(\det(\bbTheta))+\frac{1}{R}\sum_{r=1}^R \tr(\bbx_r\bbx_r^T\bbTheta)\nonumber\\
    &=-\log(\det(\bbTheta)) + \tr(\hbSigma\bbTheta).
\end{align} 
    \item The term $\rho \|\bbS\|_0$ accounts for the fact of $\bbS$ being a GSO (hence, sparse), with $\rho > 0$ being a regularization parameter that determines the desired level of sparsity in the graph.
    \item 
Finally, the equality constraint serves to embody (AS2). 
It is important to highlight that the polynomial relationship between $\bbSigma$ and $\bbS$, as implied by (AS2), is typically addressed in estimation and optimization problems through either: i) extracting the eigenvectors of $\bbSigma$ and enforcing them to be the eigenvectors of $\bbS$ \cite{ segarra2017network}, or ii) imposing the constraint $\bbSigma\bbS=\bbS\bbSigma$ \cite{buciulea2022learning}.
In contrast, our approach encodes the polynomial relation implied by (AS2) by enforcing commutativity between $\bbTheta$ and $\bbS$.
Note that if $\bbSigma$ and $\bbS$ are full-rank matrices and commute, it follows that $\bbTheta$ and $\bbS$ also commute. In other words, $\bbTheta = \bbSigma^{-1}$ can be represented as a polynomial in $\bbS$, which can be verified by the Cayley-Hamilton theorem.
\end{itemize}

It is important to note that the assumption of stationarity may seem stringent since it implies commutativity between $\bbS$ and $\bbTheta$. However, it provides more degrees of freedom than many existing methods. For example, in partial correlation methods, $\bbS$ is restricted to be $\bbS=\bbTheta$, while in our case, assuming stationarity allows $\bbTheta$ to be \emph{any} polynomial in $\bbS$. This leads to a more general approach, including partial correlation as a particular case. To further illustrate this point, consider the sparse structural equation model $\bbX=\bbA\bbX+\bbW$, with $\bbW$ being white noise. GL will identify $\bbS=\bbTheta=(\bbI_N-\bbA)^2$, while PGL will identify $\bbS=\bbA$. 

To better understand the features of PGL, let us briefly discuss the main differences relative to its two closest competitors: GSR and GL.

GSR handles Problem 1 without considering the Gaussian assumption in (AS1). As a result, the first two terms in \eqref{E:alg_GGSR}, which are associated with the log-likelihood function, are not present. This reduces the problem to inferring a sparse graph under the stationarity constraint. The stationarity assumption is incorporated into \eqref{E:alg_GGSR} through the expression $\bbTheta\bbS = \bbS \bbTheta$, which is equivalent to $\bbSigma\bbS = \bbS \bbSigma$ if $\bbSigma$ is a full-rank matrix. This property enables learning the graph by solving the following optimization problem with the commutativity constraint between $\hbSigma$ and $\bbS$:

\begin{alignat}{1}\label{E:alg_GSR}
    \!\!&\! \underset{\bbS \in \ccalS} {\mathsf{minimize}}\quad
     \rho \|\bbS\|_0  \; \; \; \mathsf{subject~to}  \; \; \; \hbSigma\bbS=\bbS\hbSigma, 
\end{alignat}
where the constraint is typically relaxed as $\|\hbSigma\bbS-\bbS\hbSigma\|_F^2\leq \epsilon$ to account for the fact that we have $\hbSigma\approx \bbSigma$.
By assuming stationarity, the formulation in \eqref{E:alg_GSR} allows the sample covariance to be modeled as any polynomial in $\bbS$, making it more general than the formulation in \eqref{E:alg_GGSR}. However, the absence of Gaussianity in \eqref{E:alg_GSR} means that it is no longer a maximum likelihood estimation. As a result, correctly identifying the ground truth $\bbS$ requires very reliable estimates of $\hbSigma$, which usually entails having access to a large number of signals to set $\epsilon$ close to zero. This is indeed a challenge, especially in setups with a large number of nodes.

For the second scenario, suppose we simplify (AS2) and instead of considering $\bbTheta$ as any polynomial in $\bbS$, we restrict it to a particular case with the following structure: $\bbTheta = \bbSigma^{-1} = \sigma\bbI+\delta\bbS$. Then, up to the diagonal values and scaling issues, the sparse matrix $\bbS$ to be estimated and $\bbTheta=\bbSigma^{-1}$ are the same. Consequently, it suffices to optimize over one of them, leading to the well-known GL formulation:
\begin{alignat}{2}\label{E:glasso} 
      \!\!&\! \underset{\bbTheta\succeq 0, \bbTheta \in \ccalS_{\bbTheta}} {\mathsf{minimize}} \
      && -\log(\det(\bbTheta)) + \tr(\hbSigma\bbTheta) + \rho \|\bbTheta\|_0. 
\end{alignat} 
The main advantages of \eqref{E:glasso} relative to \eqref{E:alg_GGSR} are that the number of variables is smaller and the resulting problem (after relaxing the $\ell_0$ norm) is convex. 
The main drawback is that by forcing the support of $\bbS$ and $\bbTheta$ to be the same, the set of feasible graphs (and their topological properties) is more limited. Indeed, GL can only estimate graphs that are positive definite, while the problem in \eqref{E:alg_GGSR} can yield any symmetric matrix.
Remarkably, when the model assumed in  \eqref{E:glasso} holds true (i.e., data is Gaussian and $\bbTheta$ is sparse), GL is able to find reliable estimates of $\bbS$ even when the number of samples $R$ is fairly low. 
On the other hand, simulations will show that GL does a poor job estimating $\bbS$ when the relation between the precision matrix and $\ccalG$ is more involved. 

\vspace{0.1cm}

In conclusion, from a conceptual point of view, our formulation reaches a favorable balance between GL and graph-stationarity approaches. This leads to the following two main advantages i) a more general model than GL since our approach models $\bbTheta$ as any polynomial in $\bbS$ and ii) a model with more structure than the graph-stationarity approaches due to the incorporation of (AS1).
However, it is important to note that the optimization in \eqref{E:alg_GGSR}, even if the $\ell_0$ norm is relaxed, lacks convexity due to the presence of a bilinear constraint that couples the optimization variables $\bbTheta$ and $\bbS$. 
These challenges will be addressed in the subsequent section.


\section{Biconvex relation and algorithm design}\label{sec:biconvexrelax_PGL_AlgDesign}

As explained in the previous section, the problem in \eqref{E:alg_GGSR} is not convex and this challenges designing an algorithm to find a good solution. This section reformulates \eqref{E:alg_GGSR}, develops an iterative algorithm, referred to as PGL, to estimate $\bbS$ and $\bbTheta$, and characterizes its convergence to a coordinate-wise minimum point. The proposed approach involves several modifications: 1) we replace the $\ell_0$-norm with an elastic net regularizer, which is convex \cite{zou2005regularization}; and 2) we relax the commutativity constraint using an inequality instead of an equality. Next, we explain step by step the resulting formulation.

\subsection{Biconvex relaxation}

The first modification to reformulate \eqref{E:alg_GGSR} is to relax the constraint that imposes commutativity between $\bbS$ and $\bbTheta$. Such a constraint is stringent and significantly narrows the feasible solution set of \eqref{E:alg_GGSR}, which may not be practical in real-world scenarios. Furthermore, considering that our access to the covariance (or precision) matrix is limited to its sampled estimates, enforcing exact commutativity is excessively restrictive. To mitigate this, we relax the original constraint by replacing the matrix equality $\bbTheta\bbS = \bbS\bbTheta$ with the scalar Frobenius norm-based inequality $\|\bbTheta\bbS - \bbS\bbTheta\|_F^2 \leq \delta$. This modification not only expands the feasible region but also endows the model with a greater degree of robustness.

The second modification addresses the non-convexity of the objective in \eqref{E:alg_GGSR}, which originates from the use of the $\ell_0$-norm. To alleviate this issue, we relax the problem using an elastic net regularizer. Specifically, we replace the $\rho \|\bbS\|_0$ penalty with $\rho\big(\|\bbS \|_1 + \frac{\eta}{2\rho} \|\bbS \|_F^2\big)$, where the parameter $\eta$ controls the trade-off between the $\ell_1$-norm and the Frobenius norm components, and is typically set to a very small value. Although elastic net regularizers have demonstrated practical effectiveness, alternative methods for relaxing the $\ell_0$-norm exist (see, for example, \cite{tibshirani1996regression,candes2008enhancing}), each offering distinct trade-offs in computational complexity, convergence speed, and theoretical underpinnings.

With the incorporation of these two modifications, we reformulate the original graph learning problem presented in \eqref{E:alg_GGSR} as follows
\begin{equation}\label{E:alg_GGSR_no_relax}
\begin{array}{rll}
&\underset{\bbTheta \succeq 0, \bbS \in \ccalS}{\mathsf{minimize}} & -\log(\det(\bbTheta)) + \tr(\hbSigma\bbTheta) + \rho \|\bbS \|_1 + \frac{\eta}{2} \|\bbS \|_F^2, \\
&\mathsf{subject~to} & \|\bbTheta\bbS - \bbS\bbTheta \|_F \leq \delta,
\end{array}
\end{equation}
In this setup, $\delta$ serves as a hyperparameter chosen according to the quality of the estimation of $\hbSigma$ which affects the estimation of $\bbTheta$. A smaller value of $\delta$ is appropriate when the quality of $\hbSigma$ is high, which typically corresponds to having a sample size $R$ that is substantially larger than the number of nodes. While the relaxation of the commutativity constraint enhances the robustness of our formulation to data quality and simplifies the optimization by reducing the number of Lagrange multipliers, the product of $\bbTheta$ and $\bbS$ still introduces nonconvexity into the problem. The way we propose for dealing with the (updated) biconvex constraint is to solve \eqref{E:alg_GGSR_no_relax} using an alternating optimization algorithm. This family of algorithms is widely used to approximate nonconvex problems by dividing the original problem into several convex subproblems and solving them with respect to each of the variables by fixing all the others. In our particular case, this methodology involves alternately solving for $\bbTheta$ with $\bbS$ held fixed, and then updating $\bbS$ using the newly updated $\bbTheta$, at each iteration.


%

In the subsequent two subsections, we delve into the detailed methodologies employed to solve each of the two subproblems. Following this, we outline the overall algorithm and discuss its convergence properties. To simplify exposition, in the remainder of the section, we will assume that $\bbS$ represents the \emph{adjacency} matrix of an undirected graph. Consequently, the feasible solution set for $\bbS$ is defined as:
\begin{equation}\label{E:s_set}
\ccalS \! := \! \big\{\bbS \! \in \! \reals^{N \times N} | \bbS \! = \! \bbS^{T}; \ \bbS \! \geq \! \mathbf{0}; \ \mathrm{diag}(\bbS) \! = \! \bb0; \ \bbS\bbone \! \geq \! \bbone \big\},
\end{equation}
where $\bbS$ is constrained to be symmetric with zero diagonal entries and non-negative off-diagonal elements. The additional condition $\bbS\mathbf{1} \geq \mathbf{1}$ is imposed to preclude the trivial solution, i.e., $\bbS=\bb0$. Nonetheless, the techniques presented next can be readily extended to accommodate different forms of $\bbS$.

\subsection[Solving subproblem for S]{Solving subproblem for $\bbS$}
We begin by addressing the subproblem with respect to $\bbS$, while holding $\bbTheta$ fixed. The subproblem is formulated as:
\begin{equation}\label{E:alg_GGSR_S}
\begin{array}{rll}
&\underset{\bbS \in \ccalS}{\mathsf{minimize}} & \rho\|\bbS\|_1  + \frac{\eta}{2}\|\bbS\|_F^2, \\
&\mathsf{subject~to} & \| \bbTheta\bbS - \bbS\bbTheta \|_F \leq \delta.
\end{array}
\end{equation}
To solve \eqref{E:alg_GGSR_S}, we adopt a linearized ADMM approach, which introduces an auxiliary variable $\bbP$ and leads to the following equivalent formulation:
\begin{equation}\label{E:alg_GGSR_S_admm}
\begin{array}{rll}
&\underset{\bbS \in \ccalS, \bbP }{\mathsf{minimize}} & \rho\|\bbS\|_1  + \frac{\eta}{2}\|\bbS\|_F^2, \\
&\mathsf{subject~to} & \bbTheta\bbS - \bbS\bbTheta = \bbP, \|\bbP\|_F \leq \delta.
\end{array}
\end{equation}
The augmented Lagrangian associated with \eqref{E:alg_GGSR_S_admm} is then given by
\begin{alignat}{2}\label{E:GGSR_aug_lag} 
    \!\!&\! L(\bbS,\bbP,\bbZ) = \rho\|\bbS\|_1 + && \frac{\eta}{2}\|\bbS\|_F^2 + \langle \bbZ,\bbTheta\bbS-\bbS\bbTheta - \bbP \rangle  \nonumber \\ 
    \!\!&\! && + \frac{\beta}{2}\|\bbTheta\bbS-\bbS\bbTheta-\bbP\|_F^2,
\end{alignat} 
where $\bbZ$ is the Lagrange multiplier.

To update $\bbS$ at the $t$-th iteration, we address the following minimization problem:
\begin{equation}\label{E:GGSR_S_1} 
\underset{\bbS \in \ccalS}{\mathsf{minimize}} \quad \rho\|\bbS\|_1  + \frac{\eta}{2}\|\bbS\|_F^2 +  \frac{\beta}{2}\| \bbTheta\bbS - \bbS\bbTheta - \bbP + \frac{1}{\beta}\bbZ \|_F^2,
\end{equation}
where, for the sake of simplicity, we omit the iteration subscript from $\bbTheta^{(t)}$ and $\bbZ^{(t)}$. Problem \eqref{E:GGSR_S_1} does not admit a closed-form solution due to the term $\frac{1}{2} \| \bbTheta\bbS - \bbS\bbTheta - \bbP + \frac{1}{\beta}\bbZ \|_F^2$. To deal conveniently with this problem we resort to the majorization-minimization (MM) algorithm \cite{SunBabuPalomar2017}. We denote this term as $g(\bbS)$ and proceed to majorize both $g(\bbS)$ and $\frac{\eta}{2}\|\bbS\|_F^2$ at the point $\bbS^{(t)}$, resulting in the following problem:
\begin{equation}\label{E:GGSR_S_2} 
\begin{split}
    \underset{\bbS \in \ccalS}{\mathsf{minimize}} \quad & \langle \rho\bbI_{N \times N} + \eta\bbS^{(t)} + \beta \nabla g(\bbS^{(t)}), \bbS-\bbS^{(t)} \rangle  \\ 
   & \qquad + \frac{L_1}{2}\|\bbS-\bbS^{(t)}\|_F^2,
\end{split}
\end{equation}
where $\nabla g(\bbS)$ represents the gradient of $g(\bbS)$, detailed in the equation:
\begin{alignat}{2}\label{E:GGSR_S_grad} 
    \!\!&\! \nabla g(\bbS) = \;\; && \bbTheta \bbTheta\bbS + \bbS\bbTheta\bbTheta - 2\bbTheta\bbS\bbTheta + \bbP\bbTheta\nonumber \\
   \!\!&\! && -\bbTheta\bbP+\frac{1}{\beta}(\bbTheta\bbZ-\bbZ\bbTheta).
\end{alignat}

Now, Problem \eqref{E:GGSR_S_2} has a closed-form solution, allowing for the update of $\bbS^{(t+1)}$ as follows:
\begin{equation}\label{E:GGSR_S_sol} 
     \bbS^{(t+1)} = \ccalP_{\ccalS} \Big(\bbS^{(t)} - \frac{1}{L_1}\big(\rho\bbI_{N} + \eta\bbS^{(t)} + \beta \nabla g\bigl(\bbS^{(t)}\bigl)\big)\Big),
\end{equation}
where $\ccalP_{\ccalS}$ is the projection onto the set $\mathcal{S}$ with respect to the Frobenius norm, which can be computed efficiently by the Dykstra's projection algorithm \cite{boyle1986method}. More specifically, the set $\mathcal{S}$ can be written as the intersection of two closed convex sets as follows:
\begin{equation}
\mathcal{S} = \mathcal{S}_A \cap \mathcal{S}_B,
\end{equation}
where $\mathcal{S}_A:= \left\lbrace  \mathbf{S} \in \mathbb{R}^{N \times N} \, | \, \mathbf{S} = \mathbf{S}^T \right\rbrace$ and $\mathcal{S}_B:= \left\lbrace  \mathbf{S} \in \mathbb{R}^{N \times N} \, | \, \mathbf{S} \geq 0; \, \mathrm{diag} (\mathbf{S}) = \mathbf{0}; \, \mathbf{S} \mathbf{1} \geq \mathbf{1} \right\rbrace$.
We employ Dykstra's projection algorithm \cite{boyle1986method} to compute the nearest point projection of a given point onto the intersection of sets $\mathcal{S}_A$ and $\mathcal{S}_B$. Dykstra's algorithm achieves this by alternately projecting the point onto $\mathcal{S}_A$ and $\mathcal{S}_B$ until the solution is reached. For a more comprehensive understanding of Dykstra's projection algorithm, the reader is directed to \cite{boyle1986method}. Detailed descriptions of the projection computations onto sets $\mathcal{S}_A$ and $\mathcal{S}_B$ are provided in Appendix~\ref{sec-computation-projection}.

\RestyleAlgo{ruled}

\begin{algorithm}[tb]
    \small
    \SetKwInOut{Output}{Outputs}
    \SetKwInput{KwData}{Input}
    \KwData{$\hbTheta^{(k)}$, $\hbS^{(k)}$, $\hbP^{(k)}$, $\hbZ^{(k)}$, $\rho$, $\eta$, $\beta$, $\delta$}
    \vspace{0.1cm}
    \Output{$\hbS^{(k+1)}$, $\hbP^{(k+1)}$, $\hbZ^{(k+1)}$}
    \SetAlgoLined
    \vspace{0.1cm}

    Initialize $\bbS^{(0)} \!=\! \hbS^{(k)}$, $\bbP^{(0)} \!=\! \bbP^{(k)}$, $\bbZ^{(0)} \!=\! \bbZ^{(k)}$   \\
    
    \For{$t=0$ \KwTo $T-1$}{
    \vspace{0.1cm}
    Update $\bbS^{(t+1)}$ by \eqref{E:GGSR_S_sol}; 
    \\ \vspace{0.1cm}
    Update $\bbP^{(t+1)}$ by \eqref{E:GGSR_P_sol}; 
    \\ \vspace{0.1cm}
    Update $\bbZ^{(t+1)}$ by \eqref{E:GGSR_Z}; 
    \\ \vspace{0.1cm}
    }
    \vspace{0.1cm}
    $\hbS^{(k+1)} = \bbS^{(T)}$, $\hbP^{(k+1)} = \bbP^{(T)}$, $\hbZ^{(k+1)} = \bbZ^{(T)}$.\\
    
    \caption{Inner loop for $\bbS$ update.} 
    \label{A:BSUM_S}
    
\end{algorithm}

Returning to the augmented Lagrangian in \eqref{E:GGSR_aug_lag}, we update $\bbP$ at the $t$-th iteration by solving the following problem: 
\begin{equation}\label{E:GGSR_P}
\begin{array}{rll}
&\underset{\bbP}{\mathsf{minimize}} & \frac{\beta}{2}\|\bbP-\bbTheta\bbS+\bbS\bbTheta-\frac{1}{\beta}\bbZ\|_F^2, \\
&\mathsf{subject~to} & \|\bbP\|_F \leq \delta,
\end{array}
\end{equation}
where we have simplified the notation by omitting the iteration subscripts from $\bbS^{(t+1)}$ and $\bbZ^{(t)}$. Problem \eqref{E:GGSR_P} has a closed-form solution. As a result, $\bbP^{(t+1)}$ can be updated by 
\begin{equation}\label{E:GGSR_P_sol} 
     \bbP^{(t+1)} = \ccalP_{\delta} \left(\bbTheta\bbS - \bbS \bbTheta + \frac{1}{\beta}\bbZ \right),
\end{equation}
where $\ccalP_\delta$ denotes the projection defined by: 
\begin{alignat}{2}\label{E:GGSR_P_proj} 
    \ccalP_\delta(\bbA) = \left\{ \begin{array}{ll}
\frac{\delta}{\|\bbA\|_F}\bbA & \mbox{if $\|\bbA\|_F>\delta$} \\
\bbA & \mbox{otherwise.}
\end{array}
\right.
\end{alignat}

Finally, the dual variable $\bbZ$ is updated according to:
\begin{equation}\label{E:GGSR_Z} 
    \bbZ^{(t+1)} = \bbZ^{(t)} + \beta(\bbTheta\bbS -\bbS\bbTheta-\bbP),
\end{equation}
where the iteration subscripts from $\bbS^{(t+1)}$ and $\bbP^{(t+1)}$ have been omitted for simplicity. A pseudocode of the steps to be performed for the update of $\bbS$ is summarized in Algorithm \ref{A:BSUM_S}.

If the parameter $L_1$ in \eqref{E:GGSR_S_2} is larger than the Lipschitz constant of the gradient of $\beta g\bigl(\bbS\bigl) + \frac{\eta}{2} \|\bbS\|_F^2$, then the sequence $\bigl\{ \big( \bbS^{(t)}, \bbP^{(t)} \big) \bigl\}$ converges to the optimal solution of Problem \eqref{E:alg_GGSR_S_admm}, and $\bigl\{ \bbZ^{(t)} \bigl\}$ converges to the optimal solution of the dual of problem \eqref{E:alg_GGSR_S_admm}, which follows from the existing convergence result of majorized ADMM \cite{li2016majorized}. To enhance empirical convergence rates, adopting a more proactive strategy for selecting the parameter $L_1$ is beneficial. For example, utilizing a backtracking line search to determine the stepsize in \eqref{E:GGSR_S_sol} can help to accelerate convergence.

We note that the choice of the penalty parameter $\beta$ can affect the convergence speed of the ADMM algorithm. A poorly chosen $\beta$ may lead to very slow convergence in practice. Adaptive schemes that adjust $\beta$ have been shown to often result in better practical performance. For example, we can adopt the adaptive update rule presented in \cite{boyd2011distributed}:
\begin{align}\label{E:adaptive}
\beta^{(t+1)} = \left\{
\begin{array}{ll}
  \tau^{\mathrm{inc}} \beta^{(t)}, & \text{if } \big \| \mathbf{r}^{(t)} \big\|_F > \mu \big \| \mathbf{s}^{(t)} \big\|_F, \\
\beta^{(t)}/\tau^{\mathrm{dec}}, & \text{if } \big \| \mathbf{s}^{(t)} \big\|_F > \mu \big \| \mathbf{r}^{(t)} \big\|_F, \\
  \beta^{(t)}, & \text{otherwise},
\end{array}
\right.
\end{align}
where $\mu >1$, $\tau^{\mathrm{inc}} > 1$, and $\tau^{\mathrm{dec}} > 1$ are predefined parameters. Here, $\mathbf{r}^{(t)}$ and $\mathbf{s}^{(t)}$ represent the primal and dual residuals at iteration $t$, respectively. They are defined as
\begin{equation*}
    \mathbf{r}^{(t)} = \bbTheta \bbS^{(t)} - \bbS^{(t)} \bbTheta - \mathbf{P}^{(t)},
\end{equation*}
and 
\begin{equation*}
    \mathbf{s}^{(t)} = \beta^{(t)} \bbTheta \big( \mathbf{P}^{(t)} - \mathbf{P}^{(t-1)}\big) -\beta^{(t)} \big( \mathbf{P}^{(t)} - \mathbf{P}^{(t-1)}\big) \bbTheta.
\end{equation*}
Although it can be challenging to prove the convergence of ADMM when $\beta$ varies by iteration, the convergence theory established for a fixed $\beta$ remains applicable if one assumes that $\beta$ becomes constant after a finite number of iterations.

\subsection[Solving subproblem for ]{Solving subproblem for $\bbTheta$}
Using the formulation from \eqref{E:alg_GGSR_no_relax}, we now turn our attention to the subproblem for $\bbTheta$
\begin{equation}\label{E:alg_GGSR_theta}
\begin{array}{rll}
&\underset{\bbTheta \succeq 0}{\mathsf{minimize}} & -\log(\det(\bbTheta)) + \tr(\hbSigma\bbTheta), \\
&\mathsf{subject~to} & \| \bbS \bbTheta - \bbTheta \bbS \|_F \leq \delta.
\end{array}
\end{equation}
Similarly to the approach taken for the $\bbS$ subproblem, we reformulate the subproblem \eqref{E:alg_GGSR_theta} for $\bbTheta$ as follows
\begin{equation}\label{E:GGSR_theta_lin}
\begin{array}{rll}
&\underset{\bbTheta \succeq 0, \ \bbQ}{\mathsf{minimize}} & -\log(\det(\bbTheta)) + \tr(\hbSigma\bbTheta), \\
&\mathsf{subject~to} & \bbS\bbTheta - \bbTheta\bbS = \bbQ, \ \|\bbQ\|_F \leq \delta.
\end{array}
\end{equation}
The augmented Lagrangian associated with \eqref{E:GGSR_theta_lin} is given by
\begin{alignat}{2}\label{E:GGSR_theta_AL} 
    L(\bbTheta,\bbQ,\bbY) = & -\log(\det(\bbTheta)) + \tr(\hbSigma\bbTheta) 
       \\ 
    & + \frac{\beta}{2}\|\bbS \bbTheta-\bbTheta\bbS-\bbQ + \frac{1}{\beta}\bbY\|_F^2. \nonumber
\end{alignat} 
To update $\bbTheta$ at the $t$-th iteration, we solve the following optimization problem
\begin{alignat}{2}\label{E:GGSR_Theta_1} 
    \underset{\bbTheta \succeq 0}{\mathsf{minimize}} \ & -\log(\det(\bbTheta)) + \tr(\hbSigma\bbTheta) \nonumber \\
    & \qquad + \frac{\beta}{2}\|\bbS \bbTheta-\bbTheta\bbS -\bbQ + \frac{1}{\beta}\bbY \|_F^2,
\end{alignat}
where we have omitted the iteration subscripts from $\bbQ^{(t)}$ and $\bbY^{(t)}$ for simplicity.

Let $f(\bbTheta) = \frac{1}{2}\|\bbS \bbTheta-\bbTheta\bbS -\bbQ + \frac{1}{\beta}\bbY \|_F^2$. We then construct the majorizer of the objective function in \eqref{E:GGSR_Theta_1} at the point $\bbTheta^{(t)}$ and obtain
\begin{alignat}{2}\label{E:GGSR_Theta_1_maj} 
    \underset{\bbTheta \succeq 0}{\mathsf{minimize}} \ & -\log(\det(\bbTheta)) + \langle \beta\nabla f(\bbTheta^{(t)}) + \hbSigma, \bbTheta-\bbTheta^{(t)}\rangle \nonumber \\
    & \qquad + \frac{L_2}{2}\|\bbTheta-\bbTheta^{(t)}\|_F^2,
\end{alignat}
where $\nabla f(\bbTheta)$ denotes the gradient of $f(\bbTheta)$
\begin{alignat}{2}\label{E:GGSR_Theta_grad} 
     \nabla f(\bbTheta) & =  \bbS \bbS \bbTheta + \bbTheta\bbS \bbS - 2\bbS \bbTheta \bbS  \nonumber \\
& \quad + \bbQ\bbS -\bbS\bbQ+\frac{1}{\beta}(\bbS \bbY-\bbY\bbS).
\end{alignat}

\RestyleAlgo{ruled}

\begin{algorithm}[tb]
    \small
    \SetKwInOut{Output}{Outputs}
    \SetKwInput{KwData}{Input}
    \KwData{$\bbSigma$, $\hbTheta^{(k)}$, $\hbS^{(k+1)}$, $\hbQ^{(k)}$, $\hbY^{(k)}$, $\beta$, $\delta$}
    \vspace{0.1cm}
    \Output{$\hbTheta^{(k+1)}$, $\hbQ^{(k+1)}$, $\hbY^{(k+1)}$}
    \SetAlgoLined
    \vspace{0.1cm}

    Initialize $\bbTheta^{(0)} = \hbTheta^{(k)}$, $\bbQ^{(0)} =  \hbQ^{(k)}$, $\bbY^{(0)} = \hbY^{(k)}$\\
    
    \For{$t=0$ \KwTo $T-1$}{
    \vspace{0.1cm}
    Update $\bbTheta^{(t+1)}$ by \eqref{E:GGSR_Theta_sol}; 
    \\ \vspace{0.1cm}
    Update $\bbQ^{(t+1)}$ by \eqref{E:GGSR_Q_sol}; 
    \\ \vspace{0.1cm}
    Update $\bbY^{(t+1)}$ by \eqref{E:GGSR_Y}; 
    \\ \vspace{0.1cm}
    }
    \vspace{0.1cm}
    $\hbTheta^{(k+1)} = \bbTheta^{(T)}$, $\hbQ^{(k+1)} = \bbQ^{(T)}$, $\hbY^{(k+1)} = \bbT^{(T)}$.\\
    
    \caption{Inner loop for $\bbTheta$ update.} 
    \label{A:BSUM_T}
\end{algorithm}

\noindent \textbf{Lemma 1.} \textit{The optimal solution of problem \eqref{E:GGSR_Theta_1} is \cite{de2022learning} }
\begin{alignat}{2}\label{E:GGSR_Theta_sol} 
     \bbTheta^{(t+1)} = \bbU \left( \frac{\bbLambda+ \sqrt{\bbLambda^2+ \frac{4}{L_2}}}{2}\right)\bbU^{\top},
\end{alignat}
where $\bbLambda$ and $\bbU$ contain the eigenvalues and eigenvectors of $\bbTheta^{(t)}-\frac{1}{L_2}\left(\beta\nabla f(\bbTheta^{(t)})+ \hbSigma\right)$, respectively.

Then, $\bbQ$ is updated by addressing the following problem
\begin{equation}\label{E:GGSR_Q}
\begin{array}{rll}
&\underset{\bbQ}{\mathsf{minimize}} & \frac{\beta}{2}\|\bbQ-\bbS \bbTheta+\bbTheta\bbS -\frac{1}{\beta}\bbY\|_F^2, \\
&\mathsf{subject~to} & \|\bbQ\|_F \leq \delta,
\end{array}
\end{equation}
where the iteration subscripts from $\bbTheta^{(t+1)}$ and $\bbY^{(t)}$ have been omitted for simplicity. Similar to the case of updating $\bbP$, we update $\bbQ$ as 
\begin{alignat}{2}\label{E:GGSR_Q_sol} 
     \bbQ^{(t+1)} = \ccalP_{\delta} \left(\bbS \bbTheta - \bbTheta\bbS + \frac{1}{\beta}\bbY\right).
\end{alignat}
Finally, the Lagrange multiplier $\bbY$ is updated as follows
\begin{alignat}{2}\label{E:GGSR_Y} 
    \bbY^{(t+1)} = \bbY^{(t)} + \beta(\bbS \bbTheta -\bbTheta\bbS-\bbQ),
\end{alignat}
where the iteration subscripts from $\bbTheta^{(t+1)}$ and $\bbQ^{(t+1)}$ have been similarly omitted for clarity.

We can also use the
adaptive strategy in \eqref{E:adaptive} to adjust $\beta$ during iterations, with $\mathbf{r}^{(t)}$ and $\mathbf{s}^{(t)}$ defined as
\begin{equation*}
    \mathbf{r}^{(t)} = \bbS \bbTheta^{(t)} - \bbTheta^{(t)} \bbS -  \mathbf{Q}^{(t)},
\end{equation*}
and 
\begin{equation*}
    \mathbf{s}^{(t)} = \beta^{(t)} \bbS \big( \mathbf{Q}^{(t)} - \mathbf{Q}^{(t-1)}\big) -\beta^{(t)}\big( \mathbf{Q}^{(t)} - \mathbf{Q}^{(t-1)}\big) \bbS.
\end{equation*}
A pseudocode of the steps to be performed for the update of $\bbTheta$ is summarized in Algorithm \ref{A:BSUM_T}.

\subsection{Graph-learning algorithm and convergence analysis}
Leveraging the results presented in the previous two subsections, the steps to run our iterative scheme to find a solution $(\hbTheta,\hbS)$ to \eqref{E:alg_GGSR_no_relax} are summarized in Algorithm \ref{A:BSUM}.

Before presenting the associated theoretical analysis, several comments regarding the implementation of Algorithm \ref{A:BSUM} are in order:
\begin{itemize}
\item For simplicity, the algorithm considers a fixed number of iterations, but a prudent approach is to monitor the cost reduction at each iteration and implement an early exit approach if no meaningful improvement is achieved.

\item The value of the hyperparameters $\rho$, $\eta$ and $\delta$ is an input to the algorithm. We note that $\eta$ is typically set to a small value to guarantee that the (sparsity promoting) $\ell_1$ norm plays a more prominent role. Moreover, the value of $\delta$ should be chosen based on the quality of the estimate of the sample covariance matrix $\hbSigma$. The higher the number of observations $R$ (hence, the better the quality of $\hbSigma$), the smaller the value of $\delta$. Similarly, if the number of nodes $N$ is high, the value of $\delta$ should be re-scaled accordingly, so that the constraint does not become too restrictive.

\item The update of $\bbS$ poses the primary computational challenge, mainly due to the complex nature of its estimation. In contrast to $\bbTheta$, which is primarily estimated from the data matrix $\hbSigma$, the estimation of $\bbS$ relies on its interplay with $\bbTheta$ through the relaxed commutativity constraint. This indirect relationship adds to the complexity of the estimation, as it does not directly benefit from data-driven insights, often necessitating greater precision in solving the corresponding subproblem. Furthermore, the constraints imposed on $\bbS$ are substantially more complex than those on $\bbTheta$, thereby increasing the computational load to obtain a solution that lies within the feasible set. To alleviate the computational demand, we may employ a relatively loose stopping criterion for the $\bbTheta$ subproblem, which can expedite convergence without significantly affecting the quality of the solution.

\end{itemize}

To establish the theoretical convergence of Algorithm~\ref{A:BSUM}, we begin by introducing several definitions and (mild) assumptions pertinent to Problem~\eqref{E:alg_GGSR_no_relax}. 

Let $f(\bbTheta, \bbS)$ denote the objective function and $\mathcal{X}$ the feasible set of Problem \eqref{E:alg_GGSR_no_relax}, respectively. We define $(\bar{\bbTheta}, \bar{\bbS})$ as a \textit{block coordinatewise minimizer} of Problem \eqref{E:alg_GGSR_no_relax} if: 
\begin{equation}
    f ( \bar{\bbTheta}, \bar{\bbS} ) \leq f(\bbTheta, \bar{\bbS}), \quad \forall \, \bbTheta \in \bar{\mathcal{X}}_{\bbTheta},
\end{equation}
and 
\begin{equation}
    f ( \bar{\bbTheta}, \bar{\bbS} ) \leq f(\bar{\bbTheta}, \bbS), \quad \forall \, \bbS \in \bar{\mathcal{X}}_{\bbS},
\end{equation}
where $\bar{\mathcal{X}}_{\bbTheta} = \{ \bbTheta \in \mathbb{R}^{N \times N} \, | \, (\bbTheta, \bar{\bbS}) \in \mathcal{X} \}$, and $\bar{\mathcal{X}}_{\bbS} = \{ \bbS \in \mathbb{R}^{N \times N} \, | \, (\bar{\bbTheta}, \bbS) \in \mathcal{X}\}$.

\RestyleAlgo{ruled}

\begin{algorithm}[tb]
    \small
    \SetKwInOut{Output}{Outputs}
    \SetKwInput{KwData}{Input}
    \KwData{$\hbSigma$, $\rho$, $\eta$, $\beta$, and $\delta$}
    \vspace{0.1cm}
    \Output{$\hbTheta$ and $\hbS$}
    \SetAlgoLined
    \vspace{0.1cm}

    Initialize $\hbTheta^{(0)} = \hbSigma^{-1}$ \\
    Initialize $\hbS^{(0)}$ by solving \eqref{E:alg_GSR} \\ \vspace{0.1cm}
    Initialize $\bbP^{(0)}$, $\bbQ^{(0)}$, $\bbY^{(0)}$, and $\bbZ^{(0)}$ to zero    
    
    \For{$k=0$ \KwTo $K-1$}{

    \vspace{0.1cm}
    Update \! $\hbS^{(k+1)}$ \! by running Algorithm \ref{A:BSUM_S}; 
    \\ \vspace{0.1cm}

    
    Update \! $\hbTheta^{(k+1)}$ \! by running Algorithm \ref{A:BSUM_T}; 
    \\ \vspace{0.1cm}


    
    }
    \vspace{0.1cm}
    $\hbTheta=\hbTheta^{(K)}$,  $\hbS=\hbS^{(K)}$.\\
    
    \caption{Polynomial Graphical Lasso (PGL) } 
    \label{A:BSUM}
\end{algorithm}

Furthermore, we introduce the following assumptions for our analysis:
\begin{assumption}\label{assumption1}
    The parameter $\delta$ in Problem~\eqref{E:alg_GGSR_no_relax} is sufficiently large to ensure that the feasible set of the subproblem~\eqref{E:alg_GGSR_S} is nonempty at every iteration.
\end{assumption}
We require Assumption~\ref{assumption1} because the intersection $\mathcal{S} \cap \{ \bbS \in \mathbb{R}^{N \times N} | , \| \bbTheta\bbS - \bbS\bbTheta \|_F \leq \delta \}$ may otherwise be empty, implying that the feasible set of subproblem~\eqref{E:alg_GGSR_S} could be nonexistent. However, Assumption~\ref{assumption1} is relatively mild, as we can always choose a sufficiently large $\delta$ to ensure that the feasible set of subproblem~\eqref{E:alg_GGSR_S} remains nonempty at every iteration. Given Assumption~\ref{assumption1}, our algorithm is guaranteed to find a minimizer of subproblem~\eqref{E:alg_GGSR_S} throughout its iterations. Additionally, the feasibility of subproblem~\eqref{E:alg_GGSR_theta} is inherently assured.

\begin{assumption}\label{assumption2}
    The matrix $\hbSigma$ in Problem~\eqref{E:alg_GGSR_no_relax} is positive definite.
\end{assumption}
Assumption~\ref{assumption2}, which requires all the eigenvalues to be nonzero, guarantees that subproblem \eqref{E:alg_GGSR_theta} is well defined. Without this assumption, the objective function in \eqref{E:alg_GGSR_theta} may fail to achieve a finite minimum value. In cases where Assumption~\ref{assumption2} may not hold, incorporating an norm regularizer for $\bbTheta$ would bound the solution, thereby ensuring the existence of a minimizer.

\begin{figure*}[t]
	\centering
	
	\begin{minipage}[c]{.35\textwidth} 
		\includegraphics[width=\textwidth]{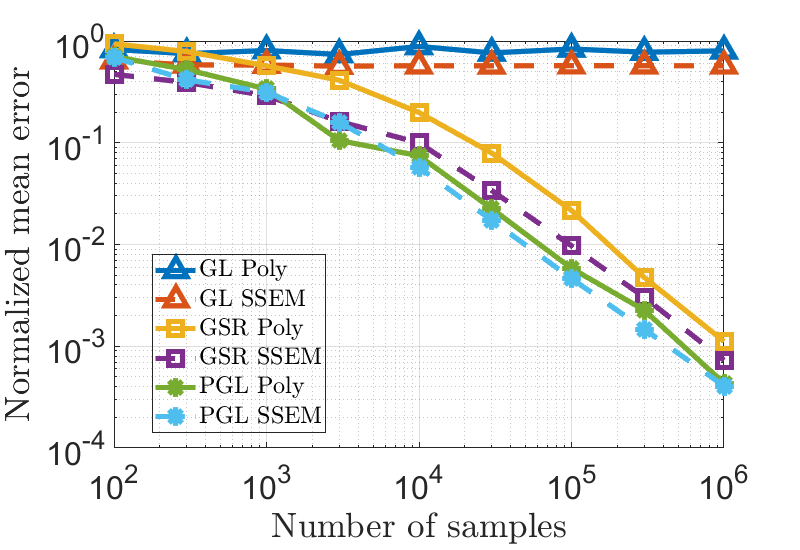}
		\centering{\small (a)}
	\end{minipage}%
	\begin{minipage}[c]{.35\textwidth} 
		\includegraphics[width=\textwidth]{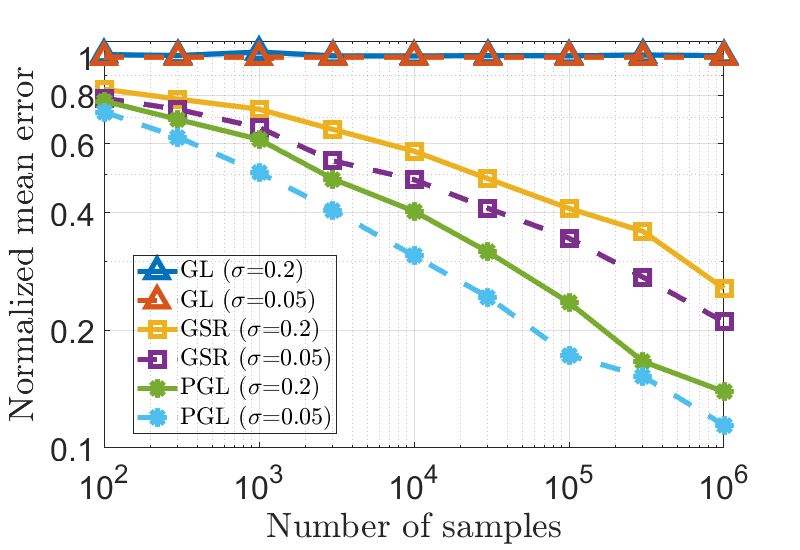}
		\centering{\small (b)}
	\end{minipage}%
	\begin{minipage}{.35\textwidth} 
		\includegraphics[width=\textwidth]{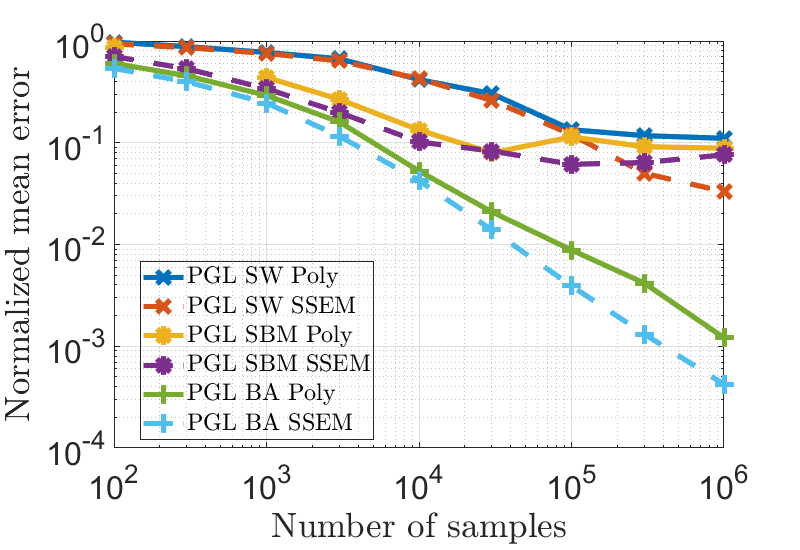}
		\centering{\small (c)}
	\end{minipage}
	\caption{Graph estimation error $\mathrm{nme}(\bbS^*,\hbS)$ vs. number of samples $R$ for different graph learning approaches (PGL, GL, and GSR) and simulations setups. Specifically, the six lines reported in each subplot correspond to the combination of (a) 3 different graph learning methods and 2 different covariance setups (SSEM and Poly) for a noise-free scenario; (b) 3 different graph learning schemes and 2 noise levels $\sigma\in\{0.05,0.2\}$ for a Poly setup; and (c) 3 graph generation models and 2 covariance models (SSEM and Poly) for the PGL algorithm in a noise-free scenario.}	
	\label{F:exp_1}
\end{figure*}

%
\begin{theorem}\label{thm:convergence}
    Let $\big\{\big(\hbTheta^{(k)}, \hbS^{(k)} \big) \big\}_{k \in \mathbb{N}} $ be a sequence generated by Algorithm~\ref{A:BSUM}. Under Assumptions~\ref{assumption1} and \ref{assumption2}, every limit point of $\big\{\big(\hbTheta^{(k)}, \hbS^{(k)} \big) \big\}_{k \in \mathbb{N}}$ is a block coordinatewise minimizer of Problem \eqref{E:alg_GGSR_no_relax}.
\end{theorem}

The proof of Theorem~\ref{thm:convergence} is deferred to Appendix~\ref{sec-convergence-proof}. This theorem asserts the subsequence convergence of our algorithm to a \textit{block coordinatewise minimizer} of Problem \eqref{E:alg_GGSR_no_relax}. When the \textit{block coordinatewise minimizer} lies within the interior of the feasible set, it becomes a stationary point. The theorem's assertions are significant both theoretically and practically. As discussed earlier in the paper, GMRFs are a particular case of Gaussian graph-stationary processes. Taking this into account, we can always initialize our algorithm using the solution estimated by GL (which is optimal for GMRF) and then, run iteratively the updates over $\bbTheta$ and $\bbS$ in Algorithm~\ref{A:BSUM}, to get an (enhanced) coordinatewise minimum estimate.

\section{Numerical experiments}\label{S:numerical_experiments}

This section evaluates quantitatively the performance of PGL. Since PGL can be understood as a generalization of the widely adopted GL (with $\bbTheta$ being any polynomial in $\bbS$), in most test cases, we will test both algorithms. Similarly, we also test the learning performance of GSR \cite{segarra2017network}, which assumes stationarity but not Gaussianity. The performance results obtained from both synthetic and real-data experiments are summarized in Figs. \ref{F:exp_1}-\ref{F:exp_7}. Unless otherwise stated, to assess the quality of the estimated GSO, we use the normalized mean error between the estimated and true $\bbS$. Mathematically, this entails computing\footnote{Results for other recovery metrics (including accuracy and F1 score) as well as additional simulations can be found both in our conference precursor \cite{buciulea2023learning} and in our online repository \url{https://github.com/andreibuciulea/GaussSt_TopoID}. }
\begin{equation}\label{eq:nerr}
    \mathrm{nme}(\bbS^{*}, \hbS) = \frac{\| \bbS^{*} - \hbS\|_F^2}{\| \bbS^{*} \|_F^2},
\end{equation}
where $\hbS$ and $\bbS^*$ represent the estimated and true $\bbS$ respectively. Moreover, for the synthetic experiments we test the graph learning algorithms over $100$ realizations of random graphs $\{\ccalG_i\}_{i=1}^{100}$ and report the average normalized mean error $\frac{1}{100}\sum_{i=1}^{100}\mathrm{nme}(\bbS_i^{*}, \hbS_i)$.

If not specified otherwise, in our synthetic experiments, we consider graphs with $N=20$ nodes generated using the Erdös-Rényi (ER) model with a link probability of $p=0.1$. Regarding the generation of the graph signals, three different setups for the covariance matrices have been studied.
For the setup referred to as ``Poly'', the covariance matrix $\bbSigma$ is generated as a random polynomial of the GSO of the form $\bbSigma=(\sum_{l=0}^{L-1} h_l \bbS^l)^2$, where $h_l$ are random coefficients drawn from a normalized zero-mean Gaussian distribution, and the square operator guarantees matrix $\bbSigma$ to be positive definite.
The setup referred to as ``SSEM'' constructs the covariance matrices following the sparse structural equation model~\cite{BazerqueGeneNetworks} for graph signal generation as $\bbSigma = (\bbI-\bbS)^2$, where $\bbS$ is selected to ensure that $\bbSigma$ is positive definite.
The setup referred to as ``MRF'' constructs the covariance matrices following the assumptions made by GL as $\bbSigma=(\mu\bbI+\nu\bbS)^{-1}$, where $\mu$ is some positive number large enough to assure that $\bbSigma^{-1}$ is positive definite and $\nu$ is some positive random number.

\subsection{Test case 1: Estimation error vs. number of samples for multiple synthetic scenarios.}

In this first test case, we employ synthetic scenarios for testing the performance of our approach in terms of  nme$(\bbS^*,\hbS)$ vs. $R$ and also compare the results with other methods from the literature. The different scenarios considered are detailed below.

\vspace{0.1cm}
\noindent \textbf{Error vs. number of samples for different graph learning methods and signal models.}
The results of the experiment depicted in Fig. \ref{F:exp_1}a, compare the nme$(\bbS^*,\hbS)$ (y-axis) of various algorithms with respect to the number of available samples $R$ (x-axis). 
Moreover, we consider SSEM and Poly setups for data generation.  
The results shown in Fig. \ref{F:exp_1}a reveal that: i) PGL outperforms its competitors, ii) the error decreases as $R$ increases, and iii) estimation is more accurate for SSEM than for Poly. Next, we discuss these three main findings in greater detail. All algorithms do a better job estimating the graph for the SSEM model. Since $\bbTheta$ for SSEM is a specific second-order polynomial in $\bbS$ it can be seen as a particular case of Poly, and consequently, a scenario from which the graph structure is easier to estimate. Indeed, while GL fails to estimate the graph for the Poly model, it is able to estimate some of the links for the SSEM. 
However, the estimation error of GL is quite large and does not change with the number of samples $R$, demonstrating that the poor performance is due to a model mismatch. This will be further confirmed in Section \ref{sec:sims_test_case_noisyMRF}, where we simulate a GMRF data generation setup that aligns perfectly with the assumptions made by GL. Regarding PGL and GSR, we observe that the error decreases almost linearly with the number of samples $R$. Perhaps more importantly, we also observe that, as $R$ increases, the gap between PGL and GSR diminishes. For example, while for the Poly case GSR needs 10 times more samples than PGL to achieve an error of $10^{-1}$, GSR only needs 3 times more samples than PGL to achieve an error of $10^{-3}$. This illustrates that, as expected, the gains associated with assuming Gaussianity are stronger when the number of observations $R$ is small, vanishing as $R$ grows very large. 

\vspace{0.1cm}
\noindent \textbf{Error vs. number of samples for noisy observations.}
Next, we assess the performance of the graph-learning algorithms in the presence of additive white Gaussian observation noise. The results are shown in Fig. \ref{F:exp_1}b. As in the previous test case, we report nme$(\bbS^*,\hbS)$ vs $R$ for PGL, GL and GSR. The difference here is that we consider only the more intricate signal generation model (Poly) and two normalized noise levels ($\sigma$ = 0.05, $\sigma$ = 0.2). The main observations in this case are: i) PGL outperforms GSR and GL, ii) the error for PGL and GSR decreases as $R$ increases, while the one for GL is flat and close to $1$,  iii) the estimation performance for PGL and GSR worsens as the noise level $\sigma$ increases, deteriorating noticeably with respect to the setup in Fig. \ref{F:exp_1}a, and iv) the gap between PGL and GSR grows.   
The findings in i) and ii) are consistent with those found in the previous test case. Finding iii) is expected and common in all graph-learning approaches. Finally, the larger gap in finding iv) is due to the fact that this is a more challenging scenario (high-order polynomial covariances and observation noise), and, as a result, the higher level of sophistication of PGL relative to GSR translates into more noticeable gains. 


\vspace{0.1cm}
\noindent \textbf{Error vs. number of samples for different graph models.}
This test case considers network models other than ER. In particular, three different types of graphs are considered: 1) Small World (SW) graphs with mean node degree $4$ and rewiring probability $0.15$; 2) Stochastic block model (SBM) graphs with $4$ clusters, and intra and inter-cluster edge probability of 0.8 and 0.05, respectively; and 3) Barabási-Albert (BA) graphs with $2$ edges to attach at every step. As in the first test case, we consider two types of signal generation models: SSEM and Poly. Fig. \ref{F:exp_1}c reports the error vs. the number of samples for the six combinations considered (3 types of graphs and 2 types of signal generation models). 
The results show that there is a significant difference in performance between SW, SBM, and BA, which is part due to the sparsity level present in each graph. One of the assumptions codified in our model is that the graph is sparse, and, as a result, our algorithm does a better job estimating BA (the one with the lowest average degree) than SBM and SW (the one with the highest average degree). Finally, we also note that the estimation error achieved with SSEM consistently outperforms that of the Poly setup. These findings align with the results presented in Fig. \ref{F:exp_1}a for ER graphs and are in accordance with the theoretical discussion that postulated SSEM as a specific instance of Poly.

\subsection{Test case 2: Noisy GMRF graph signals.}\label{sec:sims_test_case_noisyMRF}
In this test case, the goal is to assess the behavior of PGL for GMRF observations, which is the setup that motivated the development of the GL algorithm.

\begin{figure}
\begin{minipage}[c]{.45\textwidth}
	\includegraphics[width=\textwidth]{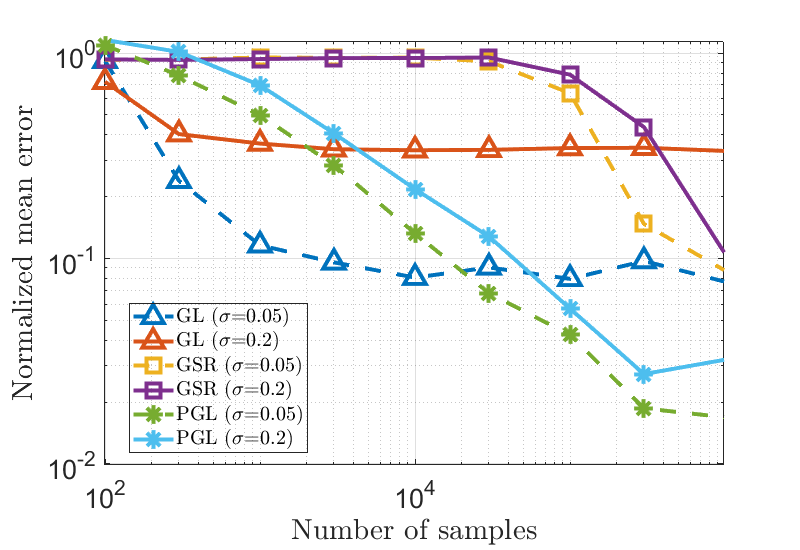}
 \end{minipage}%
 \caption{Graph estimation error $\mathrm{nme}(\bbS^*,\hbS)$ vs. number of samples $R$ for data generated according to a GMRF. We consider three different graph learning schemes (PGL, GL, and GSR) and two levels of additive white noise ($sigma\in\{0.05,0.2\}$), giving rise to the six lines in the figure.}
	\label{F:exp_4}
\end{figure}

\vspace{0.1cm}

\begin{table*}
\centering

\begin{tabular}{lllllllll}
\hline \hline
Alg. $\backslash$ $N$       & \;\; 20 & \;\; 30 & \;\;  40 & \;\;  50 & \;\;  60 & \;\; 70 & \;\; 80 & \;\;  Metric    \\ \hline \hline
\vspace{0.1cm}
\textbf{PGL-CVX}      & $2.37 \cdot 10^{1}$ & \;\; $3.88 \cdot 10^{1}$ & \;\; $1.29 \cdot 10^{2}$ & \;\; $1.11\cdot 10^{3}$ & \;\; $3.13\cdot 10^{3}$  & \;\; ----  & \;\; ----  & \;\; \multirow{2}{*}{\textbf{Time} (s)} \\ 
\vspace{0.1cm}
\textbf{PGL-Alg.\ref{A:BSUM}}          & $2.49\cdot 10^0$ & \;\; $2.72\cdot 10^0$ & \;\; $2.88\cdot 10^0$ & \;\; $3.81\cdot 10^0$ & \;\; $4.45\cdot 10^0$ & \;\; $5.86\cdot 10^0$ & \;\; $6.75\cdot 10^0$ \\  \hline
\vspace{0.1cm}
\textbf{PGL-CVX}          & $7.98\cdot 10^{-4}$ & \;\; $3.12 \cdot 10^{-3}$ & \;\; $1.35 \cdot 10^{-2}$ & \;\;$2.59 \cdot 10^{-2}$ & \;\; $9.16\cdot 10^{-2}$  & \;\; ----  & \;\; ----  & \;\; \multirow{2}{*}{nme$(\bbS^*,\hbS)$} \\ 
\vspace{0.1cm}
\textbf{PGL-Alg.\ref{A:BSUM}}         & $4.66\cdot 10^{-4}$ & \;\; $1.30\cdot 10^{-3}$ & \;\; $2.18 \cdot 10^{-3}$ & \;\;$5.07 \cdot 10^{-3}$ & \;\; $1.59 \cdot 10^{-2}$ & \;\; $1.43 \cdot 10^{-1}$ & \;\; $7.94 \cdot 10^{-1}$ \\ \hline
\end{tabular}
\vspace{0.1cm}
\caption{Values of the running time and $\mathrm{nme}(\bbS^*,\hbS)$ for two different implementations of PGL:one using an off-the-shelf convex solver (PGL-CVX) and the other one using the efficient method in Algorithm \ref{A:BSUM} (PGL-Alg.\ref{A:BSUM}). The results are shown for different graph sizes $N$ and $-$ indicates that the running time was more than 2 hours. 
}\label{Tab:1}
\end{table*}

\noindent \textbf{Estimation error considering noisy GMFR signals.} 
In this experiment, we replicate the scenario from Fig.\ref{F:exp_1}b, utilizing a GMFR model to generate the signals. The performance of PGL, GL and GSR for two different noise levels, $\sigma \in \{0.05, 0.2\}$, is depicted in Fig. \ref{F:exp_4}. The main observations are: i)
across all considered approaches, increasing the noise level $\sigma$ leads to a deterioration in terms of  nme$(\bbS^*,\hbS)$; ii) GSR always performs worse than PGL, providing very poor results when $R<10^{4}$; and iii) GL outperforms PGL when the number of observations $R$ is small. Findings i) and ii) are unsurprising and consistent with the behavior observed in the previous experiments, showcasing the benefits of considering the log-likelihood regularization in the optimization run by PGL. Regarding iii), GL outperforming PGL is expected, since the latter needs to "learn" the particular polynomial between $\bbTheta$ and $\bbS$.  

However, as the value of $R$ increases, the error in PGL gradually decreases, while the error in GL remains relatively constant, leading to PGL outperforming GL for large values of $R$. This behavior is more surprising and can be attributed to the fact that GL focuses on learning the precision matrix $\bbTheta$, while PGL balances the accuracy in terms of both the precision $\bbTheta$ and the graph $\bbS$. Since the reported error focuses on the estimation $\bbS$, the values of the diagonal of $\bbTheta$ are not relevant for nme$(\bbS^*,\hbS)$ and this can explain that GL, which is a maximum likelihood estimate for $\bbTheta$, does not yield the minimum nme$(\bbS^*,\hbS)$.

\subsection{Test case 3: Computational complexity.}
Here, we compare the runtime obtained by the efficient implementation of our PGL scheme provided in Algorithm \ref{A:BSUM} and a generic block-coordinate alternating minimization algorithm that uses CVX \cite{grant2014cvx}, the most common off-the-shelf solver for convex problems. 

\vspace{0.1cm}
\noindent \textbf{Computational complexity and estimation error.} 
The objective of this experiment is to evaluate the running time and estimation error for different versions of our algorithm as the number of nodes increases.
In particular, the experiment focuses on the Poly setup, utilizing $R = 10^6$ graph signals, and averages the results over 50 graph realizations. Table \ref{Tab:1} lists the elapsed time required to obtain the graph and precision estimates for problems with different numbers of nodes $N$ using two algorithms: 1) solving the optimization in \eqref{E:alg_GGSR_no_relax} with a block coordinate approach where the minimization over $\bbTheta$ given $\bbS$  and the minimization over $\bbS$ given $\bbTheta$ are run using CVX (this algorithm is labeled as PGL-CVX) and 2) employing the efficient scheme outlined in Algorithm \ref{A:BSUM} (this choice is labeled as PGL-Alg.\ref{A:BSUM}). To guarantee that the results are comparable, the nme$(\bbS^*,\hbS)$ is also reported.  
Examining the listed running times, we observe that as the number of nodes increases, both solvers require more time to estimate the graph (note that the number of variables scales with $N^2$). More importantly, there exists a noticeable difference between PGL-CVX and PGL-Alg.3. Not only the latter is faster for small graphs, but the gains grow significantly as $N$ increases. Note that the results for graphs with more than $N=60$ nodes are not reported for PGL-CVX, since the computation time exceeded two hours.
In terms of nme$(\bbS^*,\hbS)$, our approach achieves similar (slightly better) results than PGL-CVX.
In conclusion, the experiment demonstrates that the efficient implementation described in Section \ref{sec:biconvexrelax_PGL_AlgDesign} is more efficient than readily available convex solvers, rendering it particularly well-suited for large graphs while yielding comparable nme$(\bbS^*,\hbS)$.



\subsection{Test case 4: Real data scenarios.}
Finally, we compare different graph-learning algorithms (including PGL) in the context of two different graph-aware applications. The details and results are provided next.  

\begin{figure}
\begin{minipage}[c]{.45\textwidth}
	\includegraphics[width=\textwidth]{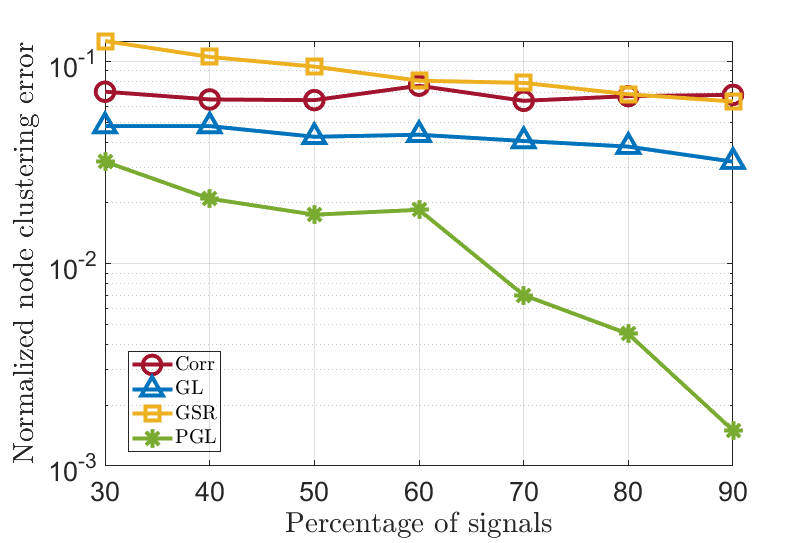}
 \end{minipage}%
 \caption{The experiment considers 40 companies of the S\&P 500 from 4 different sectors, using as nodal signals the daily returns in the period 2010-2015. Four different graph-learning schemes are considered: correlation networks (Corr), PGL, GL and GSR. After learning the graph, a spectral clustering method is implemented. The figure shows the normalized node clustering error (fraction of wrongly clustering nodes) as the percentage of available signals to learn the graph increases. }
	\label{F:exp_6}
\end{figure}

\vspace{0.1cm}
\noindent \textbf{Stock graph-based clustering from returns.} 
For this real-data experiment, we tested our graph learning algorithm on financial data and further performed a clustering task. 
Specifically, 40 companies from 4 different sectors of the S\&P 500 were selected (10 companies from each sector) and the market data (log returns) of each company in the period 2010-2015 was retrieved. This gives rise to a data matrix of size $\bbX \in \reals^{40 \times 1510}$. 
In this application, as in many others dealing with graph learning, we do not have access to the ground truth graph. Hence, we cannot quantify the quality of the estimated network directly by using a metric like $\mathrm{nme}$ or F1 score. As a result, we need to assess the quality of the estimated graph indirectly, using the graph as input for an ulterior task. In this experiment, we use the graph to estimate the community each company belongs to in an unsupervised manner. More specifically, we implement the following pipeline: 1) estimate several graphs from a subset of the available graph signals, 2) use spectral clustering to group the nodes into 4 communities (as many as sectors were selected), and 3) compute the ratio of incorrectly clustered nodes. 
To obtain more reliable results, we averaged the clustering errors over 50 realizations in which the subset of graph signals was chosen uniformly at random.
The vertical axis of Fig.\ref{F:exp_6} represents the normalized clustering error $\frac{1}{50}\sum_{i=1}^{50}\frac{N_{w}^{(i)}}{N}$ which is computed as the average of the fraction between the number of wrongly clustered nodes $N_w$ and the total number of nodes $N$ over $50$ graph realizations. The horizontal axis of Fig. \ref{F:exp_6} represents the percentages of graph signals used to estimate $\hbS$. Results are provided for 4 graph-learning approaches: PGL, GL, GSR, and ``Corr'', which estimates the graph as a thresholded version of the correlation matrix.
The idea is that companies from the same sector have stronger ties among them and, as a result, when running a graph-based clustering method, the 4 sectors should arise.
Based on the results presented in Fig.\ref{F:exp_6}, it can be observed that PGL outperforms the other alternatives and additionally, as the number of available signals increases, the clustering error for PGL drops significantly. The superiority of PGL may indicate that considering more complex relationships among stocks (beyond the simple correlations considered in ``Corr'' or the partial correlations considered in GL) is a better model to understand the dependencies between log-returns in the stock market.
On the other hand, GSR shows high clustering error with a limited number of samples, but it improves as the number of available samples increases. This observation aligns with our earlier discussion in Section \ref{S:netw_reconstruction}, where we highlighted that this particular model offers greater generality but necessitates a substantial number of samples to accurately estimate the graph.

\begin{figure}
\begin{minipage}[c]{.45\textwidth} 
	\includegraphics[width=\textwidth]{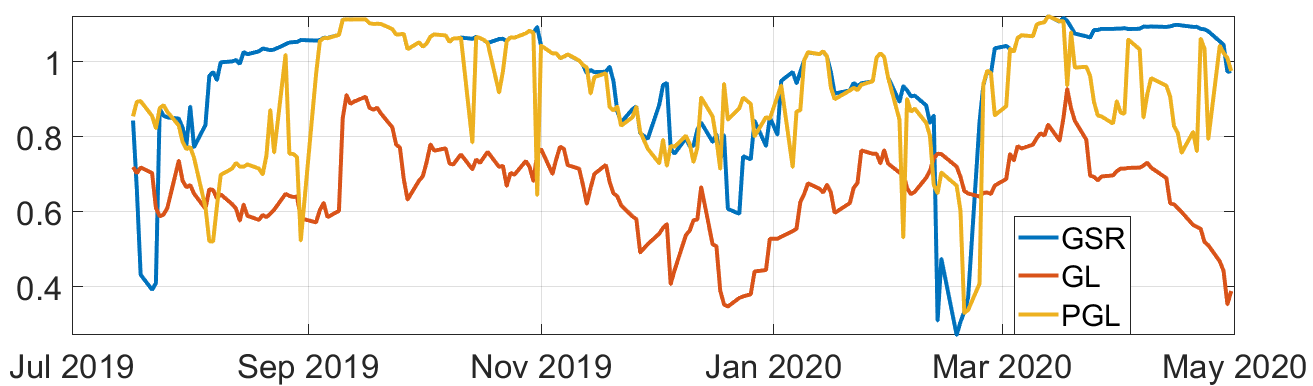}
	\centering{\small (a)}
\end{minipage}

\begin{minipage}[c]{.45\textwidth} 
    \includegraphics[width=\textwidth]{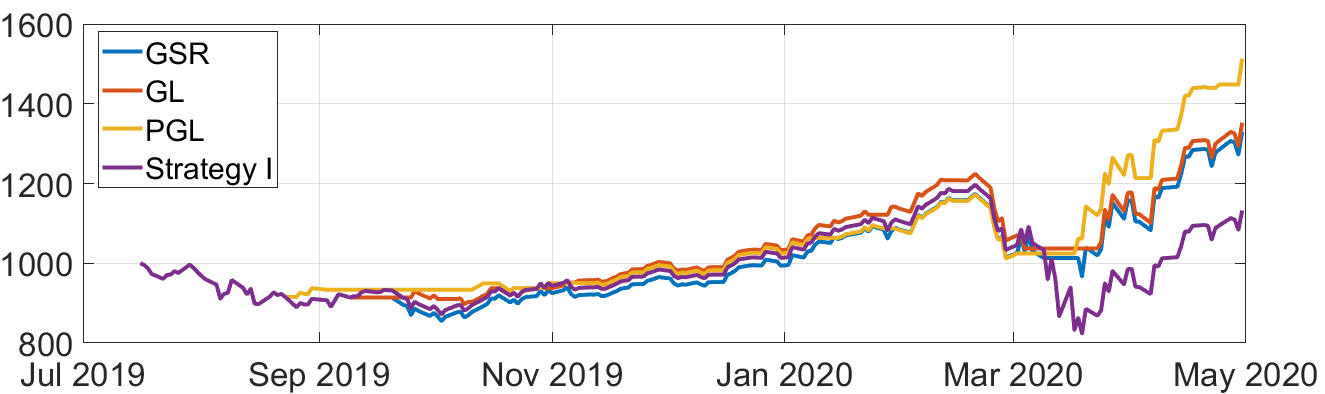}
	\centering{\small (b)}
 \end{minipage}
 \caption{This experiment learns the graph connecting the 7 FAAMUNG stocks using as signals the closing price from July 2019 to May 2020. Three different graph learning methods are used (PGL, GL, and GSR) and a different graph is learned for every day (using the signals of the previous 30 days). Subplot (a) shows the value of the algebraic connectivity ($\lambda_2$) associated with each one of the $200\times 3$ estimated graphs. Subplot (b) shows the value of the portfolio for 4 different investment strategies, 3 of which are based on the algebraic connectivity estimated in the subplot (a).}
	\label{F:exp_7}
\end{figure}

\vspace{0.1cm}
\noindent \textbf{Learning sequential graphs for investing.}
This experiment deals with a different real-world problem and dataset. We still look at stocks, but consider now the close price of the 7  FAAMUNG\footnote{Facebook, Amazon, Apple, Microsoft, Uber, Netflix, and Google} companies from Jul 2019 to May 2020. The goal is to design an investment strategy to maximize the benefits using as input a graph describing the relationships among the companies. Inspired by the approach in \cite{cardoso2020learnfin}, we first build a graph, analyze its connectivity and then, invest (or not) in a stock according to the graph connectivity. To be more specific, we use the close price to estimate multiple $7\times 7$ adjacency matrices. We estimate a total of 200 matrices, where each adjacency matrix (graph) is estimated in a rolling window fashion. 
The window consists of 30 consecutive days and for each graph estimation, we shift the window one day at a time. 
These graph estimations help to visualize how the graph topology changes during this time period. 
The finding in \cite{cardoso2020learnfin} is that big changes in the graph connectivity indicate opportunities to invest. To that end, we keep track of the algebraic connectivity value, which is the second smallest eigenvalue ($\lambda_2$) of the estimated Laplacian matrix. The lower the value, the less connected the graph is and, as a result, the easier breaking the graph into multiple components is. Fig.\ref{F:exp_7} shows the value of $\lambda_2$ for each of the 200 considered graphs (each associated with a 30-day period). Then, using the approach in \cite{cardoso2020learnfin} we invest only if $\lambda_2$ is below a fixed threshold. We learn the graph using 3 methods (GL, GSR, and PGL) and, for each of them, we select the best possible threshold (the one that maximizes the benefits). Correlation was not used here due to its poor performance.
The results of applying the graph-connectivity-based investment strategy to the graphs estimated with each of the algorithms are shown in Fig. \ref{F:exp_7}. The blue line labeled as ``Strategy I'' represents the benefits of investing every day the available amount and is used as a baseline.
By analyzing the results obtained, we can observe that: i) the graph-based strategies outperform (gain more money than) the baseline; ii) the strategies based on GL and GSR provide similar gains; and iii) the strategy based on GSSR yields the highest gains. This provides additional validation for the graph-learning methodology proposed in this paper.

\section{Conclusions}\label{S:conclusions}
This paper has introduced PGL, a novel scheme for learning a graph from nodal signals, with our key contribution being the modeling of the signals as Gaussian and stationary on the graph. This approach opens the door to a graph-learning formulation that leverages the advantages of GL (needing a relatively small number of signals to get a good estimation of the graph structure) while encompassing a more comprehensive model (because it handles cases where the precision matrix can be any polynomial form of the sought graph). Given the increased complexity and nonconvex nature of the resulting optimization problem, we have developed a low-complexity algorithm that alternates between estimating the graph and precision matrices and have characterized its convergence to a block coordinatewise minimum. To assess its efficacy, we have conducted numerical simulations comparing PGL with various alternatives, using both synthetic and real data. The results have showcased the benefits of our approach, motivating the adoption and further investigation of the proposed graph-learning methodology. 


\appendices
\section{Computations of projections}\label{sec-computation-projection}

We present the details about how to compute the projections $\mathcal{P}_{\mathcal{S}_A}$ and $\mathcal{P}_{\mathcal{S}_B}$.

The computation of $\mathcal{P}_{\mathcal{S}_A}$ is straightforward as follows:
\begin{equation}
\mathcal{P}_{\mathcal{S}_A} (\mathbf{A}) = \big(\mathbf{A} + \mathbf{A}^\top \big)/2.
\end{equation}

The projection $\mathcal{P}_{\mathcal{S}_B}$ is defined as the minimizer of the optimization problem as follows:
\begin{equation}\label{eq:pro-sc}
\mathcal{P}_{ \mathcal{S}_B} (\mathbf{A}) := \underset{\mathbf{X} \in \mathcal{S}_B}{\mathsf{arg \, min}} \, \frac{1}{2} \|\mathbf{X} - \mathbf{A} \|_{F}^2.
\end{equation}
To compute the projection $\mathcal{P}_{\mathcal{S}_B}$, we solve Problem \eqref{eq:pro-sc} row by row. For the $j$-th row, we solve the following problem,
\begin{equation}\label{P_B_row}
\begin{array}{rll}
&\underset{\mathbf{x} \in \mathbb{R}^N}{\mathsf{minimize}} & \frac{1}{2}\| \mathbf{x} - \mathbf{a} \|^2,  \\  
& \mathsf{subject~to} & \mathbf{x}^\top \mathbf{1} \geq 1, \, x_j = 0, \, \mathbf{x}_{\setminus j} \geq \mathbf{0},
\end{array}
\end{equation} 
where $\mathbf{a} \in \mathbb{R}^N$ contains all entries of the $j$-th row of $\mathbf{A}$, $x_j$ denotes the $j$-th entry of $\mathbf{x}$, and $\mathbf{x}_{\setminus j} \in \mathbb{R}^{N-1}$ contains all entries of $\mathbf{x}$ except the $j$-th one. 

Let $\hat{\mathbf{x}}$ denote the optimal solution of Problem \eqref{P_B_row}. Proposition \ref{solution-PB} below, proved in Appendix~\ref{sec-solution-PB}, presents the optimal solution of Problem \eqref{P_B_row}.

\begin{proposition}\label{solution-PB}
The optimal solution $\hat{\mathbf{x}}$ to Problem \eqref{P_B_row} can be obtained as follows:
\begin{itemize}
\item If $\sum_{i \neq j} \max (a_i, \, 0) \geq 1$, then $\hat{x}_j = 0$, and $\hat{x}_i = \max (a_i, \, 0)$, for $i \neq j$.

\item If $\sum_{i \neq j} \max (a_i, \, 0) < 1$, then $\hat{x}_j = 0$, and $\hat{x}_i = \max (a_i + \phi, \, 0)$, for $i \neq j$, where $\phi$ satisfies $\sum_{i \neq j} \max (a_i + \phi, \, 0) = 1$.
\end{itemize}
\end{proposition} 

Several efficient approaches have been developed to tackle the piecewise linear equation $\sum_{i \neq j} \max (a_i + \phi, 0) = 1$. Among these, the sorting-based method described in \cite{condat2016fast} is noteworthy. Central to this method is the sorting of the vector $\mathbf{a}$, which constitutes the most computationally intensive step, generally necessitating $\mathcal{O}(N \log N)$ operations.

\section{Proof of Proposition \ref{solution-PB}}\label{sec-solution-PB}
\begin{proof}
The Lagrangian of the optimization in \eqref{P_B_row} is
\begin{equation*}
L (\mathbf{x}, u, \mathbf{v}) = \frac{1}{2}\| \mathbf{x} - \mathbf{a} \|^2 - u \big(\mathbf{x}^\top \mathbf{1} - 1 \big) - \langle \mathbf{v}_{\setminus j},\, \mathbf{x}_{\setminus j}\rangle + v_j x_j,
\end{equation*}
where $u \in \mathbb{R}$ and $\mathbf{v} \in \mathbb{R}^N$ are KKT multipliers. Let $(\hat{\mathbf{x}}, \hat{u}, \hat{\mathbf{v}} )$ be the primal and dual optimal point. Then $(\hat{\mathbf{x}}, \hat{u}, \hat{\mathbf{v}} )$ must satisfy the KKT system:
\begin{align}
\hat{x}_i - a_i - \hat{u} - \hat{v}_i &= 0, \quad \mathrm{for} \ i \neq j; \label{TT1-kkt1}\\
\hat{x}_j - a_j - \hat{u} + \hat{v}_j &= 0; \label{TT1-kkt2}\\
\ \hat{x}_i \geq 0, \ \hat{v}_i \geq 0, \ \hat{v}_i \hat{x}_i &= 0, \quad \mathrm{for} \ i \neq j; \\
\hat{x}_j = 0, \ \hat{u} \geq 0, \ \hat{\mathbf{x}}^T \mathbf{1} & \geq 1; \label{TT1-kkt3} \\
\hat{u} \big(  \hat{\mathbf{x}}^T \mathbf{1} - 1 \big) &= 0; \label{TT1-kkt4} 
\end{align}
Therefore, for any $i \neq j$, it holds that $\hat{x}_i = a_i + \hat{u} + \hat{v}_i$. Then we obtain the following results:
\begin{itemize}
\item If $a_i + \hat{u} < 0$, then $\hat{v}_i = -a_i + \hat{u}$ and $\hat{x}_i = 0$, following from the fact that $\hat{v}_i \hat{x}_i = 0$ and $\hat{x}_i \leq 0$.

\item If $a_i + \hat{u}_r \geq 0$, then $\hat{v}_i = 0$. This can be obtained as follows: if $\hat{x}_i = 0$, then $\hat{v}_i = - (a_i + \hat{u}) \leq 0$. Since $\hat{v}_i \geq 0$, one has $\hat{v}_i = 0$; On the other hand, if $\hat{x}_i \neq 0$, then $\hat{v}_i = 0$, following from the fact that $\hat{v}_i \hat{x}_i = 0$.
As a result, we get $\hat{x}_i = a_i + \hat{u}$.
\end{itemize}
Overall, we obtain that
\begin{equation}\label{eq:solution_xi}
\hat{x}_j = 0, \quad \mathrm{and} \quad \hat{x}_i = \max (a_i + \hat{u}, \, 0), \quad \forall \, i \neq j.
\end{equation}

On the other hand, $\hat{\mathbf{x}}$ and $\hat{u}$ satisfy that $\hat{\mathbf{x}}^T \mathbf{1} \geq 1$, $\hat{u} \geq 0$, and $\hat{u} \big(  \hat{\mathbf{x}}^T \mathbf{1} - 1 \big) = 0$. To this end, we can obtain the following results:
\begin{itemize}
\item If $\sum_{i \neq j} \max (a_i, \, 0) \geq 1$, then $\hat{u} = 0$, indicating that $\hat{x}_i = \max (a_i, \, 0)$, for any $i \neq j$.

\item If $\sum_{i \neq j} \max (a_i, \, 0) < 1$, then $\hat{u} \neq 0$. This is because $\hat{u} = 0$ will result in $\hat{\mathbf{x}}^\top \mathbf{1} < 1$, which does not satisfy the KKT system. Together with the KKT condition that $\hat{u} \big(  \hat{\mathbf{x}}^T \mathbf{1} - 1 \big) = 0$, one has $\hat{\mathbf{x}}^\top \mathbf{1} = 1$. Therefore, one obtains that, for any $i \neq j$, $\hat{x}_i = \max (a_i + \hat{u}, \, 0)$, where $\hat{u}$ is chosen such that $\sum_{i \neq j} \max (a_i + \hat{u}, \, 0) = 1$.
\end{itemize}
We note that the $\phi$ in Proposition \ref{solution-PB} is exactly the dual optimal point $\hat{u}$.
\end{proof}

\section{Proof of Theorem \ref{thm:convergence}}\label{sec-convergence-proof}
\begin{proof}
The convergence result stated in Theorem \ref{thm:convergence} is based on the framework established by Theorem 2.3 in \cite{xu2013block}. To demonstrate the validity of Theorem \ref{thm:convergence}, it suffices to establish that the conditions and assumptions of Theorem 2.3 are satisfied in our context. Our approach to block updates aligns with the procedure delineated in equation (1.3a) of \cite{xu2013block}.

We first verify the conditions the requisite conditions of Theorem 2.3 in \cite{xu2013block} are met. Specifically, Theorem 2.3 stipulates that the objective function, along with the feasible set of the optimization problem, should exhibit \textit{block multiconvexity}. For Problem~\eqref{E:alg_GGSR_no_relax}, the objective function $f$ is convex with respect to each of the blocks $\bbTheta$ and $\bbS$ when the other block is fixed, a property that defines \textit{block multiconvexity} as per \cite{xu2013block}. Moreover, the function $f$ is strongly convex with respect to both $\bbTheta$ and $\bbS$.

The feasibility constraints of Problem~\eqref{E:alg_GGSR_no_relax} form a set $\mathcal{X}$ that satisfies the criteria for \textit{block multiconvexity} as defined in \cite{xu2013block}. This is due to the convexity of the individual set maps $\mathcal{X}_{\bbTheta}$ and $\mathcal{X}_{\bbS}$. The set map $\mathcal{X}_{\bbTheta}$ is defined as
\begin{equation}\label{eq:set_map_Theta}
    \mathcal{X}_{\bbTheta} = \{ \bbTheta \in \mathbb{R}^{N \times N} \, | \bbTheta \succeq \mathbf{0},  \| \, \bbTheta\bbS - \bbS\bbTheta \|_F \leq \delta \}
\end{equation}
for some given $\bbS$, and similarly, the set map $\mathcal{X}_{\bbS}$ is defined as
\begin{equation}\label{eq:set_map_S}
    \mathcal{X}_{\bbS} = \{ \bbS \in \mathbb{R}^{N \times N} \, | \, \bbS \in \mathcal{S},  \| \bbTheta\bbS - \bbS\bbTheta \|_F \leq \delta \}
\end{equation}
for some given $\bbTheta$. Consequently, the optimization subproblems with respect to $\bbTheta$ and $\bbS$ in Problem~\eqref{E:alg_GGSR_no_relax} are convex.

We now validate the assumptions required by Theorem 2.3 in \cite{xu2013block} within the context of our setting. Specifically, it is required that the objective function $f$ is bounded below over the feasible set $\mathcal{X}$, that is, $\inf_{(\bbTheta, \bbS) \in \mathcal{X}} f(\bbTheta, \bbS) > - \infty$. This is indeed the case here, because the term $\rho\|\bbS\|_1 + \frac{\eta}{2}\|\bbS\|_F^2$ is nonnegative. Additionally, the function $-\log(\det(\bbTheta)) + \tr(\hbSigma\bbTheta)$ attains a finite infimum when $\hbSigma$ is positive definite under Assumption~\ref{assumption2}. To see this, first observe that $\det(\bbTheta) \leq \| \bbTheta \|_2^N$, where $\| \bbTheta \|_2$ is the largest eigenvalue of $\bbTheta$. Consequently, $-\log\det(\bbTheta) \geq - N \log (\| \bbTheta \|_2)$. Further, since $\tr(\hbSigma\bbTheta) \geq \gamma \tr(\bbTheta) \geq \gamma \| \bbTheta \|_2$,
with $\gamma > 0$ being the smallest eigenvalue of $\hbSigma$, we obtain
\begin{equation}
    -\log\det(\bbTheta) + \tr(\hbSigma\bbTheta) \geq - N \log (\| \bbTheta \|_2) + \gamma \| \bbTheta \|_2,
\end{equation}
which indeed has a finite minimum. Thus, we conclude that $\inf_{(\bbTheta, \bbS) \in \mathcal{X}} -\log\det(\bbTheta) + \tr(\hbSigma\bbTheta) > -\infty$.

Furthermore, the existence of \textit{block coordinatewise minimizers} is assured by the compactness of the feasible set. Moreover, Theorem 2.3 in \cite{xu2013block} stipulates that set maps change continuously during iterations. Referring to \eqref{eq:set_map_Theta} and \eqref{eq:set_map_S}, it is clear that the only constraint that changes through iterations is $\|\boldsymbol{\Theta}\bbS - \bbS\boldsymbol{\Theta} \|_F \leq \delta$, while the other constraints, $\boldsymbol{\Theta} \succeq \mathbf{0}$ and $\bbS \in \mathcal{S}$, remain constant. Given that $\|\boldsymbol{\Theta}\bbS - \bbS\boldsymbol{\Theta}\|_F$ is a continuous function with respect to both $\boldsymbol{\Theta}$ and $\bbS$, the set maps indeed change continuously, satisfying the theorem's conditions.

These verifications above have demonstrated that the conditions and assumptions of Theorem 2.3 are satisfied in our context, completing the proof.
\end{proof}

\bibliographystyle{IEEEtran.bst}
\bibliography{main}

\end{document}